\pdfoutput=1
\documentclass{aa}  

\usepackage[varg]{txfonts}
\usepackage{amsmath}
\usepackage[version=4]{mhchem}

\usepackage{siunitx}

\usepackage{graphicx}
\usepackage{float}
\usepackage{multirow}
\usepackage{booktabs}
\usepackage{rotating}

\usepackage{xcolor}

\usepackage{natbib}
\bibpunct{(}{)}{;}{a}{}{,}

\renewcommand{\textcolor}[2]{%
  \ifthenelse{\equal{#1}{green}}{#2}{%
  \ifthenelse{\equal{#1}{red}}{#2}{%
  \textcolor{#1}{#2}}}}

\usepackage[pdftex, breaklinks=true, colorlinks=true, linkcolor=blue, citecolor=blue, urlcolor=blue]{hyperref}

\defcitealias{Ruaud2016}{RWH16}
\defcitealias{hasegawa1993three}{HH93}

\begin{document}

% \title{PEGASIS: A Fast, Flexible, and Rigorously Benchmarked Three-Phase Python-Based Astrochemical Code with Diffusive and Non-Diffusive Grain-Surface Processes for Modeling Interstellar Chemistry}  

\title{Role of Diffusive and Non-Diffusive Grain-Surface Processes in Cold Cores: Insights from the PEGASIS Three-Phase Astrochemical Model}

\titlerunning{PEGASIS: An Astrochemical Code for Diffusive and Non-Diffusive Grain-Surface Chemistry}

\authorrunning{Maitrey et al.}

\author{
    S. Maitrey \inst{1,2} 
     \and Liton Majumdar\inst{1,2}
     \and Varun Manilal\inst{1,2}
    \and Baibhav Srivastava \inst{1,2}
     \and Prathap Rayalacheruvu \inst{1,2}        
     \and Karen Willacy \inst{3} 
     \and Eric Herbst \inst{4}    
}

\institute{Exoplanets and Planetary Formation Group, School of Earth and Planetary Sciences, National Institute of Science Education and Research, Jatni 752050, Odisha, India \\
\email{liton@niser.ac.in; dr.liton.majumdar@gmail.com} 
    \and
    Homi Bhabha National Institute, Training School Complex, Anushaktinagar, Mumbai 400094, India 
     \and    
    Jet Propulsion Laboratory, California Institute of Technology, 4800 Oak Grove Drive, Pasadena, CA 91109, USA
    \and
    Department of Chemistry, University of Virginia, Charlottesville, VA 22904, USA
    }
             
%\date{Received 04 July 2024; Accepted 20 August 2024}

  \abstract
  % context heading (optional)
   {Cold dense cores are unique among the structures found in the interstellar medium, as they harbor a rich chemical inventory, including complex organic molecules (COMs), which will be inherited by future evolutionary stages, such as protostellar envelopes and protoplanetary disks. These molecules exist both in the gas phase and as ices accreted onto grain surfaces.} 
   % aims heading (mandatory) 
   {To model these environments, we present \textsc{Pegasis}, a new, fast, and extensible three-phase astrochemical code to explore the chemistry of cold cores, with an emphasis on the role of diffusive and non-diffusive chemistry in shaping their gas and grain chemical compositions.} 
  % methods heading (mandatory)
  {We incorporate the latest developments in interstellar chemistry modeling by utilizing the 2024 KIDA chemical network and comparing our results with current state-of-the-art astrochemical models. Using a traditional rate-equation-based approach, we implement both diffusive and non-diffusive chemistry, coupled with either an inert or chemically active ice mantle.} 
  % results
   {We identify crucial reactions that enhance the production of COMs through non-diffusive mechanisms on the grain-surface as well as the mechanisms through which they can build up in gas-phase. Across all models with non-diffusive chemistry, we observe a definite enhancement in the concentration of COMs on both grain-surface as well in the grain-mantle. Finally, our model broadly reproduces the observed abundances of multiple gas-phase species in a cold dense core TMC-1 (CP) and provides insights into its chemical age.}
  % conclusions
   {Our work demonstrates the capabilities of \textsc{Pegasis} in exploring a wide range of grain-surface chemical processes and modeling approaches for three-phase chemistry in the interstellar medium, providing robust explanations for observed abundances in cold cores, such as TMC-1 (CP). In particular, it highlights the role of non-diffusive chemistry in the production of gas-phase COMs on grain surfaces, which are subsequently chemically desorbed, especially when the precursors involved in their formation on the surfaces are heavier than atomic hydrogen.}

\maketitle
%
%________________________________________________________________

\section{Introduction} \label{sec:intro}
%% Add some motivations for chemisorption and non-diffusive chemistry
The accretion of gas-phase interstellar molecules on the cold dust grains plays a fundamental role in enriching the chemistry of molecular clouds. These grains facilitate several processes, both physical and chemical, to proceed on their surface which in turn leads to the formation of species that cannot be solely form\textcolor{green}{ed} in the gas phase \citep[see][and references therein]{tielens1982model, herbst2009complex}. \textcolor{red}{For instance, \citet{gould1963interstellar} established that dust grains must act as catalysts to account for the presence of molecular hydrogen observed in the interstellar medium.} The surface chemistry thus becomes the bridge which dictates the compositions of both the bulk ice that will exist on grains as well as the compositions of gas phase through desorption. Astrochemical models which can simulate these processes in \textcolor{green}{the} interstellar medium (and in various more evolved proto-stellar or protoplanetary environments) have existed ever since the use of radio observatories became prominent in detecting interstellar molecules \citep{agundez2013chemistry}. As the list of observed molecules keeps growing \citep[see the Cologne Database for Molecular Spectroscopy\footnote{\url{https://cdms.astro.uni-koeln.de/classic/molecules}} and][for the most recent compilation]{mcguire20222021} with more advanced observational capabilities, it has became imperative for the models to keep evolving. A good astrochemical model must be able to couple gas-phase chemistry with ice chemistry while being versatile enough to be used in \textcolor{green}{a variety of} physical environments. 

For cold cores, pioneering work by \citet{bates1951density, watson1973rate, watson1974ion, watson1976interstellar} and \citet{herbst1973formation} established the key chemical processes and pathways that paved the way to construct chemical networks. In conjunction with experimental work, sets of reactions involving detected molecules that can \textcolor{green}{occur} in the interstellar medium are identified and compiled to be simulated by a numerical scheme. Currently, the \textsc{UMIST} Database for Astrochemistry \citep{millar1991gas,millar1997umist,le2000umist,woodall2007umist,mcelroy2013umist,millar2024umist} and KInetic Database for Astrochemistry \textsc{(KIDA)} \citep{Wakelam2012,wakelam20152014,wakelam20242024} represent the most recent and well-maintained publicly available networks. The chemical networks typically encapsulate information about which chemical reactions and processes can proceed under specific physical conditions. These networks then can be fed in and used by several astrochemical codes \textcolor{red}{for modeling}. A few examples among many published astrochemical codes include the Willacy Model \citep{Willacy98}; \textsc{Alchemic} \citep{Semenov2010}; \textsc{Magickal} \citep{garrod2011formation}; GRAINOBLE \citep{taquet2012multilayer}; \textsc{MONACO} \citep{Vasyunin2013}; \textsc{astrochem}\footnote{\url{https://github.com/smaret/astrochem}} \citep{maret2013chemical}; the Rokko code \citep{Furuya2015}; the CMMC code \citep{Das2015ApJ}; \textsc{Nautilus}\footnote{\url{https://astrochem-tools.org/codes/}} \citep{Ruaud2016}; \textsc{UCLCHEM}\footnote{\url{https://uclchem.github.io/}} \citep{holdship2017uclchem}; \textsc{CHEMPL}\footnote{\url{https://github.com/fjdu/chempl}} \citep{du2021chempl}; the Acharyya Model \citep{acharyya2020effect}; and Sipil{\"a}'s Model \citep{sipila2022}. Several of these codes are now publicly available for broader use within the astrochemistry community as well. The approaches used by these codes to simulate chemical networks vary, but the rate equation method, which involves solving a system of coupled differential equations, has proven to be straightforward to implement and faster to converge \citep[see][for an overview on this and other techniques]{cuppen2013kinetic}.

Increasingly, theoretical and laboratory studies have highlighted the importance of physico-chemical processes such as photodissociation by UV radiation and cosmic rays, as well as diffusive chemistry, which extends beyond the gas phase to the grain surface and the bulk mantle \citep{andersson2008photodesorption, oberg2009formation}. This makes the inclusion of an active non-inert mantle crucial \textcolor{green}{to evaluate} the extent to which it can influence the wider gas phase chemistry. Several authors have attempted to implement this by considering each accreted ice layer distinctly \citep{taquet2012multilayer} or by considering multiple layer grouped as phases of the mantle \citep{furuya2017water} as well as extending the three-phase model of \citet{hasegawa1993three} to explore effects of bulk ice chemistry \citep{Ruaud2016,kalvans2010subsurface,garrod2013three}. 
% However, the extent to which bulk ice influences chemistry between an inert mantle as compared to an active mantle still remains uncertain.

Inclusion of ice chemistry opens up the possibility of progressive entrapment of molecules onto the grains, as the gas phase reservoir depletes which reduces both the rates of accretion and desorption. \citet{herbst2005chemistry} discusses the importance of non-thermal desorption pathways, especially, for much later ages in the total lifetime of cold cores. Works like \citet{garrod2006gas} have discussed the role of desorption pathways in reproducing abundances at much later times. Essentially, consideration of a more detailed framework of ice chemistry and how it interacts with the gas phase have suggested a late convergence with the observations to an increasing extent \citep[see][for the most recent development]{wakelam2021efficiency}. Thus, the estimation of chemical ages of these sources \textcolor{green}{is still uncertain.} 

The inner regions of dense clouds are often well-shielded, allowing the bulk mantle to persist for extended periods. However, the very low temperatures in these regions ($\sim 10\,\text{K}$) render most heavier species immobile, making diffusion-driven production of complex organic molecules (COMs) unfeasible. This contrasts with the detection of molecules such as methyl formate (\ce{HCOOCH3}) and dimethyl ether (\ce{CH3OCH3}) in colder regions \citep{cernicharo2012discovery, bacmann2012detection}. Consequently, alternative pathways have been proposed to form these COMs on grain surfaces without relying on diffusive chemistry \citep{ruaud2015modelling, chang2016unified, bergner2017complex, bergner2017methanol, shingledecker2018cosmic, jin2020formation}, with subsequent non-thermal desorption mechanisms playing a crucial role. Notably, \citet{balucani2015formation} also suggested the possibility of gas-phase formation of COMs. \citet{jin2020formation} employed a modified-rate method approach introduced in \citet{garrod2008new},
% to expand upon the experimental work of \citet{fedoseev2015experimental, fedoseev2017formation}
demonstrating that appreciable COM production can occur even in the absence of diffusive chemistry.

As mentioned earlier, astrochemical models must be adaptable and flexible to accurately simulate chemistry across various astrochemical environments by incorporating both diffusive and non-diffusive chemistry. The inclusion of non-diffusive chemical processes is crucial for modeling colder regions more accurately, whereas hotter regions require special consideration of surface chemistry as well. If surface molecules are assumed to be bound solely by weak van der Waals forces, the efficiency of any significant surface chemistry becomes negligible. \citet{acharyya2020effect} introduced the concept of chemisorbed sites, where molecules exhibit higher binding energies, thereby enabling surface chemistry even at temperatures exceeding $100\text{ K}$.

To address these challenges, we introduce \textsc{Pegasis}, a fast and flexible three-phase astrochemistry code that incorporates both diffusive and non-diffusive grain-surface processes, along with all fundamental ice chemistry mechanisms applicable across diverse astrochemical environments. In this paper, we apply \textsc{Pegasis} to investigate the role of both diffusive and non-diffusive grain-surface processes in shaping the gas-phase and grain-surface chemical compositions of cold cores, specifically targeting the Cyanopolyyne peak position (TMC-1 (CP)) in the Taurus Molecular Cloud, given its ever-expanding and chemically rich inventory. Recent detections include several oxygen-bearing complex organic molecules—such as propenal (\ce{C2H3CHO}), vinyl alcohol (\ce{C2H3OH}), methyl formate (\ce{HCOOCH3}), and dimethyl ether (\ce{CH3OCH3}) as reported by \citet{agundez2021bearing}, and ethanol (\ce{CH3CH2OH}), acetone (\ce{CH3COCH3}), and propanal (\ce{C2H5CHO}) by \citet{agundez2023detection}. Additionally, other newly identified species include 1,4-pentadiyne \cite[\ce{HCCCH2CCH};][]{fuentetaja2024discovery}, sulphur radicals \citep{cernicharo2024more}, dinitriles \citep{agundez2024rich}, polycyclic aromatic hydrocarbons (PAHs) \citep{cernicharo2024discovery,wenzel2024detection}, thioacetaldehyde \cite[\ce{CH3CHS};][]{agundez2025detection}, the 1-cyano propargyl radical \cite[\ce{HCCCHCN};][]{cabezas2025identification}, and cyclopropenethione \cite[\ce{c-C3H2S};][]{remijan2025missing}.
% In this paper, we apply \textsc{Pegasis} to investigate the role of both diffusive and non-diffusive grain-surface processes in shaping the gas and grain chemical compositions of cold cores, specifically targeting the Cyanopolyyne peak position (TMC-1 (CP)) in the Taurus Molecular Cloud\textbf{, given its ever-expanding rich chemical inventory and recent detections of several oxygen-bearing complex organic molecules---such as propenal (\ce{C2H3CHO}), vinyl alcohol (\ce{C2H3OH}), methyl
% formate (\ce{HCOOCH3}), and dimethyl ether (\ce{CH3OCH3}) reported by \citet{agundez2021bearing}, and ethanol (\ce{CH3CH2OH}), acetone (\ce{CH3COCH3}), and propanal (\ce{C2H5CHO}) by \citet{agundez2023detection}. In addition to this, other newly identified species include 1-4-pentadiyne \cite[\ce{HCCCH2CCH};][]{fuentetaja2024discovery}, sulphur radicals \citep{cernicharo2024more}, dinitriles \citep{agundez2024rich}, polycyclic aromatic hydrocarbons (PAHs) \citep{cernicharo2024discovery,wenzel2024detection}, thioacetaldehyde \cite[\ce{CH3CHS};][]{agundez2025detection}, 1-cyano propargyl radical \cite[\ce{HCCCHCN};][]{cabezas2025identification} and  cyclopropenethione \cite[\ce{c-C3H2S};][]{remijan2025missing}}.

The structure of the paper is as follows: In Section~\ref{sec:style}, we describe our model and the chemical processes it includes. Section~\ref{sec:bench} presents a benchmark comparison of \textsc{Pegasis} with the public version of \textsc{Nautilus} \citep{Ruaud2016}. In Section~\ref{sec:predict}, we examine the differences between the three-phase chemistry prescriptions of \citet{Ruaud2016} and \citet{hasegawa1993three}, and compare the predicted gas-phase abundances with observations in TMC-1 (CP), considering both diffusive and non-diffusive chemistry. While a large number of species have been detected in TMC-1, we restrict our discussion to those included in the 2024 KIDA chemical network. Additionally, we compare the ice abundances with those observed in more evolved sources to gain insights into the transformation of the chemical inventory from cold cores to later evolutionary stages. Finally, we summarize our conclusions in Section~\ref{sec:concl}.

\section{Model description}\label{sec:style}

\textsc{Pegasis} is a three-phase model (gas, grain-surface and grain-mantle)
which presents a comprehensive and flexible approach towards understanding how
molecular abundances evolve in the interstellar medium. The code has been designed to study the chemistry of molecular clouds, but it can also be easily extended to model chemistry in other evolutionary stages, such as protostars and protoplanetary disks. Following \citet{Ruaud2016}, we let chemical reactions
occur in the mantle as well as on the grain surface, albeit with a smaller rate
of diffusion. This is implemented by setting \textcolor{green}{the} diffusion barrier \textcolor{green}{for surface species} as a smaller fraction of
the binding energies \citep{Garrod2006} \textcolor{green}{than for mantle species}. The interaction between grain surface and gas-phase
chemistry occurs through accretion and desorption processes. Desorption can \textcolor{green}{occur} by thermal, non-thermal, or chemical mechanisms. For the three-phase
chemistry, we implement prescriptions from both \citet{Ruaud2016} and
\citet{hasegawa1993three}. Most of the processes are implemented as switches,
\textcolor{green}{allowing} their effects \textcolor{green}{to} be explored
\textcolor{green}{separately}. We \textcolor{green}{use} the most recent release from the KInetic Database for
Astrochemistry (KIDA) \citep{wakelam20242024} as our base chemical network. \textcolor{green}{This}
includes 7667 gas-phase reactions and 4837 grain-surface and grain-mantle
reactions and processes. \textsc{Pegasis} is \textcolor{green}{implemented} \textcolor{green}{entirely in} Python \textcolor{green}{and is} accelerated by
Numba \citep{siu_kwan_lam_2024_11642058}. \textcolor{green}{Finally, the differential equations are solved using the} the \texttt{DLSODES} solver from the \texttt{ODEPACK}
library \citep{hindmarsh1983scientific} wrapped as a Python
extension\footnote{\url{https://github.com/kmaitreys/pylsodes}}.

% \subsection{Flexibility in choosing various approaches}
% Interstellar medium is rich with denser-than-average regions filled with
% gaseous and dust particles \citep{Herbst2001}. These interstellar clouds exist
% with varying levels of density, size, temperature and ionization
% \citep{McKee1977,Snow2006}.

% To constrain the chemistry of these regions, it is therefore imperative to
% categorically consider all the physical and chemical processes that can
% influence these attributes through astrochemical modeling. Further, these
% chemical models of interstellar clouds enable the physical interpretation of
% observational data \citep{Black1977}, which is essential in the context of
% studying objects like protoplanetary disks which follow the collapse of these
% clouds \citep{Festou2004,Stahler2004,Henning2013}.

% \textsc{Pegasis} has thus been designed in such a way that it is possible to study how
% each of the implemented physical and chemical process affect the chemistry of the clouds.
% Processes like shielding by abundant molecular species, photodesorption on
% grain surfaces, background cosmic ionization rate that can trigger the formation of
% ionic chemical species from neutral ones, and so forth can all be individually enabled,
% disabled or finely controlled.

\subsection{Modeling chemical processes}

% \textcolor{red}{(LM: Texts in green below: Introduce how to write astrochemical processes into a system of differential equations, see how Ruaud et al. has introduced // green texts not needed)} 
% \textcolor{green}{Due to the chemical complexity of interstellar clouds, the molecular abundances in these clouds almost never reach a steady state \citep{Semenov2010}. Thus, \textsc{Pegasis} employs models of chemical kinetics, which are essentially a system of differential equations. We start from a set of initial conditions and evolve the system over timescales on the order of \(t \approx 10^5 - 10^6\) yr \citep{Hasegawa1992}, or on the order of a few Myr. The choice of the former, referred to as the \textit{early-time}, comes from the work of \citet{brown1990chemical,Brown1991}, which established that by this time, gas-phase chemistry remains largely unaffected by grain-surface chemistry. The \textit{late-time}, on the other hand, can be crucial for achieving agreement with observations when the gas-phase species are depleted \citep{ruffle2000new}.}
\textsc{Pegasis} employs \textcolor{green}{a} chemical kinetics-based approach to model different chemical processes, starting from a given \textcolor{green}{set of} initial conditions and evolving a system of coupled differential equations in time. The initial conditions define the physical and chemical configuration of the model, along with the starting concentrations of the molecules to be considered. In the simplest scenario, only the interstellar elemental abundances are used initially. The rate coefficients of each reaction are calculated and used to determine the total production or destruction of each species in the simulation. 

For each species considered, \textsc{Pegasis} solves \textcolor{green}{a system of} ordinary differential equations involving each of the \textcolor{green}{chemical species that may exist in the} three phases \textcolor{green}{of gas, grain surface and grain mantle.} \textcolor{green}{These equations} describe the evolution of the abundance of \textcolor{green}{species} with time and take the form
\begin{equation}\label{eq:gaskin}
    \begin{split}
        \frac{d n_{\text{g}}(p)}{dt} \bigg|_{\text{tot}} &=
        \sum_{q} \sum_{r} k_{qr} n_{\text{g}}(q) n_{\text{g}}(r) \\
        &+ k^{\text{g}}_{\text{diss}}(q) n_{\text{g}}(q) + k_{\text{des}} (p) n_{\text{s}} (p) \\
        &- k^{\text{g}}_{\text{diss}} (p) n_{\text{g}}(p) - k_{\text{acc}} (p) n_{\text{g}}(p)  \\
        &- n_{\text{g}}(p) \sum_{r} k_{pq} n_{\text{g}}(q),
    \end{split}
\end{equation}

\begin{equation}\label{eq:surfkin}
    \begin{split}
        \frac{d n_{\text{s}}(p)}{dt} \bigg|_{\text{tot}} &=
        \sum_{q} \sum_{r} k^{\text{s}}_{qr} n_{\text{s}}(q) n_{\text{s}}(r) \\
        &+ k^{\text{s}}_{\text{diss}}(q) n_{\text{s}}(q)
        + k_{\text{acc}} (p) n_{\text{g}}(p) \\
        &+ k^{\text{m}}_{\text{swap}} (p) n_{\text{m}} (p)
        + \dfrac{dn_{\text{m}} (p)}{dt} \bigg|_{\text{m} \rightarrow \text{s}} \\
        &- n_{\text{s}} (p) \sum_{q} k^{\text{s}}_{pq} n_{\text{s}} (q) \\
        &- k^{\text{s}}_{\text{diss}} (p) n_{\text{s}} (p) - k_{\text{des}} (p) n_{\text{s}} (p)\\
        &- k^{\text{s}}_{\text{swap}} (p) n_{\text{s}} (p)
        - \dfrac{d n_{\text{s}} (p)}{dt} \bigg|_{\text{s} \rightarrow \text{m}} ,
    \end{split}
\end{equation}

\begin{equation}\label{eq:mantkin}
    \begin{split}
        \frac{d n_{\text{m}}(p)}{dt} \bigg|_{\text{tot}} &=
        \sum_{q} \sum_{r} k^{\text{m}}_{qr} n_{\text{m}}(q) n_{\text{m}}(r) \\
        &+ k^{\text{m}}_{\text{diss}}(q) n_{\text{m}}(q) \\
        &+ k^{\text{s}}_{\text{swap}} (p) n_{\text{s}} (p)
        + \dfrac{dn_{\text{s}} (p)}{dt} \bigg|_{\text{s} \rightarrow \text{m}} \\
        &- n_{\text{m}} (p) \sum_{q} k^{\text{m}}_{pq} n_{\text{m}} (q) \\
        &- k^{\text{m}}_{\text{diss}} (p) n_{\text{m}} (p) \\
        &- k^{\text{m}}_{\text{swap}} (p) n_{\text{m}} (p)
        - \dfrac{d n_{\text{m}} (p)}{dt} \bigg|_{\text{m} \rightarrow \text{s}}.
    \end{split}
\end{equation}

Here \(k_{pq}\), \(k^{\text{s}}_{pq}\) and \(k^{\text{m}}_{pq}\) are the rate
coefficients in the gas phase, grain surface and grain mantle respectively between
species \(p\) and \(q\). \textcolor{green}{$k_{\text{diss}}$ is the rate for photodissociation processes, both in the gas phase and in the surface ($s$) and mantle ($m$) phases. Freeze-out ($k_{\text{acc}}$) of gaseous molecules onto the surface, and desorption ($k_{\text{des}}$) returning surface species to the gas are included.  Transfer of material from the surface to the mantle and vice versa is included with a rate $k^s_{\text{swap}}$ and $k^m_{\text{swap}}$ respectively.  Note that this swapping does not represent the
actual movement of material across ice layers \citep{hasegawa1993three,garrod2011formation, Ruaud2016}.}
For the reactions involving two reactants (bimolecular reactions), the rate
coefficient is calculated using the modified Arrhenius formula
\citep{Kooij1893}
\begin{equation}
    k = \alpha {(T / 300)}^{\beta} \exp (-\gamma / T)
\end{equation}
% Interstellar molecules can be both neutral or charged \citep{Herbst2001} which makes it
% essential to consider the mechanisms of formation of these charged species through ionization
% processes and subsequently reactions of these charged species with each other and neutral species.
For bimolecular reactions involving charged species, we follow
\citet{Wakelam2012} and implement the temperature-dependent
modification of rate coefficients using the Su-Chesnavich capture approach discussed in
\citet{woon2009quantum}. We suggest going through
\citet{Wakelam2012,wakelam20242024}
and references therein for more information on the kinds of bimolecular reactions
included in the network.
The charged gas-phase species can also react with negatively-charged dust
grains in a neutralization reaction, \textcolor{green}{with a} rate \textcolor{green}{coefficient} given by
\begin{equation}
    k = \alpha {\left(\dfrac{T}{300}\right)}^{\beta}
\end{equation}
\textcolor{green}{Here the} activation energy \(\gamma\) \textcolor{green}{is set} to zero \textcolor{green}{since} this kind of
recombination reaction is considered barrierless \citep{geppert2008dissociative}.
The 2024 KIDA network does not \textcolor{green}{include} positively-charged grains, thus
processes \textcolor{green}{such as} collisional charging of grains as discussed in
\citet{draine1987collisional} and collisions of positively and negatively
charged grains as explored in \citet{umebayashi1983densities} \textcolor{green}{are} not included
in our model.

% \subsection{Dissociation and Ionization}
% The production of charged species occurs through processes of dissociation and
% ionization which have may have different agents and mechanisms driving them
% such as cosmic rays \citep{cesarsky1978cosmic,Nath1994,Padovani2009}, X-rays
% \citep{habing1971heating,glassgold1973heating}, Lyman-\(\alpha\) photons
% \citep{bergin2003effects,Heays2017}, UV photons
% \citep{vanDishoeck1988,van2006photoprocesses,vanDishoeck2014} as well as
% cosmic-ray induced UV photons \citep{prasad1983uv,gredel1989cosmic}. In our
% model, photodissociation can take place anywhere in the gas phase, on the grain
% surface or in the grain mantle (see Figure \ref{fig:dissionizdesorp}). 
The rate coefficient for processes involving
ionization or dissociation by cosmic rays in the gas phase is given by the standard
scaling relation \citep{Semenov2010,Wakelam2012}
\begin{equation}\label{eq:rateCR}
    k_{\text{CR}} = \alpha \zeta_{\text{CR}}
\end{equation}
where \(\zeta_{\text{CR}}\) is the cosmic-ray ionization rate for molecular
hydrogen \textcolor{green}{and the value of \(\alpha\) is molecule dependent}. 

% Additional ionization from X-rays is considered by simply adding its
% contribution \textcolor{green}{to Equation} (\ref{eq:rateCR}).

% \begin{equation}
%     k_{\text{CR+XR}} = \alpha (\zeta_{\text{CR}} + \zeta_{\text{XR}})
% \end{equation}
% A more exhaustive approach towards X-ray
% ionization that takes \textcolor{green}{species-specific ionization} into account will be considered
% in a future work.

% \textcolor{green}{As well as directly causing dissociation, cosmic rays can also } electronically excite both
% molecular and atomic hydrogen, whose \textcolor{green}{subsequent} relaxation then later produces UV photons \textcolor{green}{that can}
% dissociate or ionize molecules \citep{prasad1983uv}. \textcolor{green}{This} CR-induced UV photodissociation/photoionization \textcolor{green}{can occur in} both \textcolor{green}{the} gas phase
% and \textcolor{green}{the} grain surface \textcolor{green}{with a} rate coefficient calculated as \citep{gredel1987c, gredel1989cosmic,gredel1990cosmic,sternberg1987cosmic} 
% \begin{equation}
%     k_{\text{UVCR}} = \dfrac{\alpha}{1 - \omega}
%     \dfrac{n(\ce{H2})}{(n(\ce{H}) + 2 n(\ce{H2}))} \zeta_{\text{CR}}
% \end{equation}
% where \(\omega\), the albedo of grains in the far-ultraviolet, is taken as
% 0.5 \citep{Wakelam2012}. 

Cosmic rays can electronically excite both molecular and atomic hydrogen. The subsequent relaxation of these excited species produces UV photons, which can then dissociate or ionize molecules \citep{prasad1983uv}. This CR-induced UV photodissociation/photoionization can occur in both the gas phase and on the grain surface, with a rate coefficient calculated as \citep{gredel1987c, gredel1989cosmic, gredel1990cosmic, sternberg1987cosmic} 
\begin{equation}
    k_{\text{UVCR}} = \dfrac{\alpha}{1 - \omega}
    \dfrac{n(\ce{H2})}{(n(\ce{H}) + 2 n(\ce{H2}))} \zeta_{\text{CR}}
\end{equation}
where \(\omega\), the albedo of grains in the far-ultraviolet, is taken as 0.5 \citep{Wakelam2012}.

\textcolor{green}{For photodissociation caused by interstellar or stellar UV photons, the rate coefficient is given by}
\begin{equation}\label{eq:selfshieldphotodiss}
    k_{\text{UV}} = \alpha \exp \left(-\gamma A_V \right) F_{\text{UV}}
\end{equation}
in both the gas and the grain surface. \textcolor{green}{Here, \(A_V\) is the visual extinction and \(F_{\text{UV}}\) the far-ultraviolet flux in Draine units.} Ionization of molecules on the grain surface is not considered. 

% In the interior of molecular clouds, where density gets higher, and other
% highly dense regions of interstellar medium in general, the radiation sources
% which can cause dissociation and ionization processes suffer from attenuation.
% The built up of columns of abundant gases like \ce{H2} and \ce{CO} shield
% themselves and other molecules from getting ionized. 
Instead of relying on a
full radiative transfer \textcolor{red}{for consideration of shielding from abundant gas phase species like \ce{H2} and \ce{CO}}, we adopt the approximation suggested in
\citet{lee1996photodissociation}, where the photodissociation rates are
calculated as a function of the visual extinction (\(A_V\)) and column
densities of the \ce{H2}. For \ce{CO}, we take shielding functions from
\citet{visser2009photodissociation} and for \ce{N2}, we adopt the prescription
from \citet{li2013photodissociation}.

% \subsection{Accretion and Desorption}\label{sec:accANDdes}
The gas phase and grain surface chemistry is linked through processes of
accretion (adsorption) and desorption of the molecules. Only neutral molecules
undergo adsorption and thus there is no charged surface species in the chemical
network. Dust grains are assumed to be spherical and their size and material
density can be set by the user. The rate constant for this process is given as
\citep{Semenov2010,Wakelam2012}
\begin{equation}\label{eq:accr}
    k_{\text{acc}} = \pi \eta r^2_{d} \sqrt{\dfrac{8 k_B T_{g}}{\pi m_p}} n_{d}
\end{equation}
where \(m_p\) is the atomic mass of the species \(p\) and \(\eta\) is the
sticking efficiency, which is taken as 100\% for neutral species. The 2024 KIDA
network does not contain charged species on grains, so efficiency for them is
taken to be zero. 
% However, for \ce{H} and \ce{H2}, the value of \(\eta\) is
% taken from the experimental work of \citet{matar2010gas} and
% \citet{chaabouni2012sticking}. If the accreting grain only has one monolayer,
% it assumed to have a silicate surface for which we have
% \begin{equation}
%     \eta_{\text{sil}} =
%     \begin{cases}
%         \dfrac{1 + T_g/10}{{(1 + T_g/25)}^{5/2}},                  & \text{for \ce{H}} \\ \\
%         \dfrac{19}{20}\dfrac{1 + 5 T_g/112}{{(1 + T_g/56)}^{5/2}}, &
%         \text{for \ce{H2}}
%     \end{cases}
% \end{equation}
% and if the number of monolayer of ice is larger than one, then the surface is
% assumed to made of amorphous solid water (ASW) for which we have
% \begin{equation}
%     \eta_{\text{ASW}} =
%     \begin{cases}
%         \dfrac{1 + 5 T_g/104}{{(1 + T_g/52)}^{5/2}},               & \text{for \ce{H}}  \\ \\
%         \dfrac{19}{25}\dfrac{1 + 5 T_g/174}{{(1 + T_g/87)}^{5/2}}, & \text{for \ce{H2}}
%     \end{cases}
% \end{equation}
% The final sticking efficiency in the case of single monolayer ice is given as
% \begin{equation}\label{eq:finalstick}
%     \eta = (1 - N_{\text{tot}})\eta_{\text{sil}} + N_{\text{tot}} \eta_{\text{ASW}}
% \end{equation}
% and if there are multiple layers, it is just \(\eta = \eta_{\text{ASW}}\).

% The ionizing agents that we discussed in the previous section also trigger
% desorption or evaporation of molecules from the grain surface to back into the
% gas-phase. 
In sufficiently warm regions, molecules can evaporate thermally from
the grain surfaces and rate constant for this process is given by the
first-order Polanyi-Wigner equation
\citep{katz1999molecular,herbst2005chemistry,Semenov2010}
\begin{equation}\label{eq:type15}
    k_{\text{des}} = \nu_0 (p) \exp \left(\dfrac{-E_d}{T_d}\right)
\end{equation}
where \(\nu_{0} (p)\) is the characteristic frequency of species \(p\) on the
grain surface, approximated as a harmonic oscillator relation
\begin{equation}\label{eq:charactfreq}
    \nu_{0} (p) = \sqrt{\dfrac{2 n_s E_d}{\pi^2 m_{\text{H}} m_{p}}}
\end{equation}
by \citet{tielens1987composition} and \citet{Hasegawa1992}.
Here, \(E_d\) is the adsorption energy of \ce{H}.
Unlike \citet{collings2004laboratory}, thermal evaporation in our model only
involves the topmost layer in the sense we do not consider multilayer
thermal desorption. \textcolor{red}{We also consider desorption driven by cosmic rays following \citet{hasegawa1993new}.}

% Desorption can also happen through cosmic rays on dust grains with appropriate
% geometric cross-section (we consider grains of radius less that
% $\qty{1}{\micron}$). According to \citet{léger1985desorption}, when a iron-nuclei
% cosmic rays collides with a dust grain, it can stochastically heat it to a peak
% temperature of $\qty{70}{\kelvin}$ \citep{hasegawa1993new}. During this period,
% desorption of a molecule into gas phase becomes possible and for this, the rate
% constant can be written as \citep{hasegawa1993new}
% \begin{equation}
%     k_{\text{CR,des}} = \nu_0 \exp \left(-\dfrac{E_b}{T_{\text{CR}}}\right) \dfrac{\zeta_{\ce{Fe}}\zeta_{\text{CR}}}{\zeta_{\text{CR,0}}} \Delta t
% \end{equation}
% where \(\zeta_{\ce{Fe}}\) is the \ce{Fe} cosmic ray flux as estimated by
% \citet{léger1985desorption}, \(\zeta_{\text{CR,0}} = \qty{1.3e-17}{\per \second}\)
% is the nominal interstellar \ce{H2} cosmic ray flux and \(\Delta t\) is the
% time interval between successive heatings to \(T_{\text{CR}}\)
% \citep{hasegawa1993new}.

% We also consider photo-desorption from the grain surfaces by following the
% simple approach proposed in \citet{bertin2013indirect} to consider a single
% photo-desorption yield for all molecules rather than for each individual ones
% determined from experiment. 
For photo-desorption driven by standard
interstellar UV photons, we have the rate formula
\citep{Ruaud2016,wakelam2021efficiency}
\begin{equation}\label{eq:uvdes}
    k_{\text{UV,des}} = F_{\text{UV}} S_{\text{UV}} \exp \left(-2 A_V\right) Y_\text{pd} \dfrac{4 \pi r_d^2}{N_s}
\end{equation}
and for secondary UV photons induced by cosmic rays, we have
\begin{equation}\label{eq:uvcrdes}
    k_{\text{UVCR,des}} = F_{\text{UVCR}} S_{\text{UVCR}} Y_\text{pd} \dfrac{4 \pi r_d^2}{N_s}
\end{equation}
The yield \(Y_\text{pd}\) is taken as \(10^{-4}\) molecules per photon following
\citet{andersson2008photodesorption}.~\(F\) is the strength of the UV field in
Draine units and \(S\) is the corresponding scaling factor. For standard
interstellar UV, we take \(F_{\text{UV}} = 10^8\) photons
\(\unit{\per \centi \metre \squared \per \second}\)
\citep{oberg2007photodesorption} with \(S_{\text{UV}} = 1\) and for cosmic ray
induced UV, we take \(F_{\text{UVCR}} = 10^4\) photons
\(\unit{\per \centi \metre \squared \per \second}\) \citep{shen2004cosmic}
with \(S_{\text{UVCR}} = \zeta / 1.3 \times 10^{\-17}\), \(\zeta\) being the
\ce{H2} cosmic ray ionization rate. The factor of 2 in
equation (\ref{eq:uvdes}), accounts for the extinction of UV photons
relative to the visual extinction \(A_V\) \citep{roberge1991interstellar}.

The chemical reactions proceeding on the grain surfaces can be exothermic which
in turn can trigger desorption of a molecule to gas phase. In our model, we
implement two modes of chemical desorption. The first mode is based on the work
of \citet{garrod2007non}, which is based in the Rice-Ramsperger-Kessel (RRK)
theory \citep[see][]{Robinson1996-ft}. We define \(f\) to be the probability that a given reaction results in desorption because of this exothermicity as
\begin{equation}\label{eq:garrodchemdes}
    f = \dfrac{\nu P}{\nu_s + \nu P} = \dfrac{aP}{1 + aP}
\end{equation}
where \(a = \nu/\nu_s\) is the ratio of surface-molecule bond frequency
to the frequency at which energy is lost to the grain surface. The RRK
probability \(P\) is given by
\begin{equation}
    P = {\left(1 - \dfrac{E_d}{E_{\text{reac}}}\right)}^{s-1}
\end{equation}
where \(E_d\) is the desorption energy of the molecule being desorbed to gas
phase, \(E_\text{reac}\) is the enthalpy of formation for the reaction and
\(s\) is the number of vibrational modes in the molecule-surface bond
system. For diatomic species, \(s = 2\) and for all others,
\(s = 3 \mathcal{N}/5\),
with \(\mathcal{N}\) being the total number of atoms in the desorbing molecule.

The second mode is based on the experimental work of \citet{minissale2016dust}
and its extension in \citet{riedel2023modelling}. We adopt experimental values
for the evaporation fraction for reactions involving \ce{H} and \ce{OH} (\(f =
0.25\)), \ce{O} and \ce{H} (\(f = 0.3\)) and \ce{N} and \ce{N} (\(f = 0.5\)) as
reactants from \citet{minissale2016dust}. For the rest of reactions, the
general expression of evaporation fraction is given by
\begin{equation}
    f = \exp \left(-\dfrac{\mathcal{F} E_d}{\epsilon E_{\text{reac} }}\right)
\end{equation}
Here, \(\epsilon\) is fraction of kinetic energy produced in the exothermic
reaction retained by the reaction product.~\(\mathcal{F}\) is the the degrees
of freedom of the product and is taken as \(\mathcal{F} = 3 \mathcal{N}\).
The expression for \(\epsilon\) can be written
as
\begin{equation}
    \epsilon = {\left(\dfrac{M-m}{M+m}\right)}^2
\end{equation}
where \(M\) is the effective mass of the surface component and \(m\)
is the mass of the reaction product which receives this kinetic energy.
Because the surfaces involved can vary, being covered by ices of
different species, we adopt values of effective surface mass \(M\)
from \citet{vasyunin2017formation}. We consider three possibilities:
a water ice substrate (\(M = 48\text{ u}\)),
a \ce{CO} ice substrate (\(M = 100\text{ u }\)) and finally,
bare grains where number of surface layers is less than or equal to unity
(\(M = 120\text{ u}\)). Every surface that is not covered with
water ice or is not a bare grain is assumed to be covered with \ce{CO}
ice \citep{riedel2023modelling}. Note that, in the case of just
one reaction product, this process involves two bodies (one desorbing
product and the surface component). In the reactions with two desorbing
products, we accordingly modify the expression for \(\mathcal{F}\)
accordingly to include contribution of the second product. Finally, the
total time-dependent evaporation fraction is given as \citep{riedel2023modelling}
\begin{equation}
    f (i, t) = \sum_j f_j (i) \cdot \dfrac{n^{\star}_j (t)}{n^{\star}_{\text{surf}} (t)}
\end{equation}
where \(f_j (i)\) is the individual fraction for the three surface types
(bare grain, \ce{H2O} and \ce{CO}), \(n^{\star}_j\) is the surface sites populated
by surface type \(j\) and \(n^{\star}_{\text{surf}}\) is the total surface site
abundance.
\begin{figure*}[htbp]
    \centering
    \includegraphics[width=\textwidth]{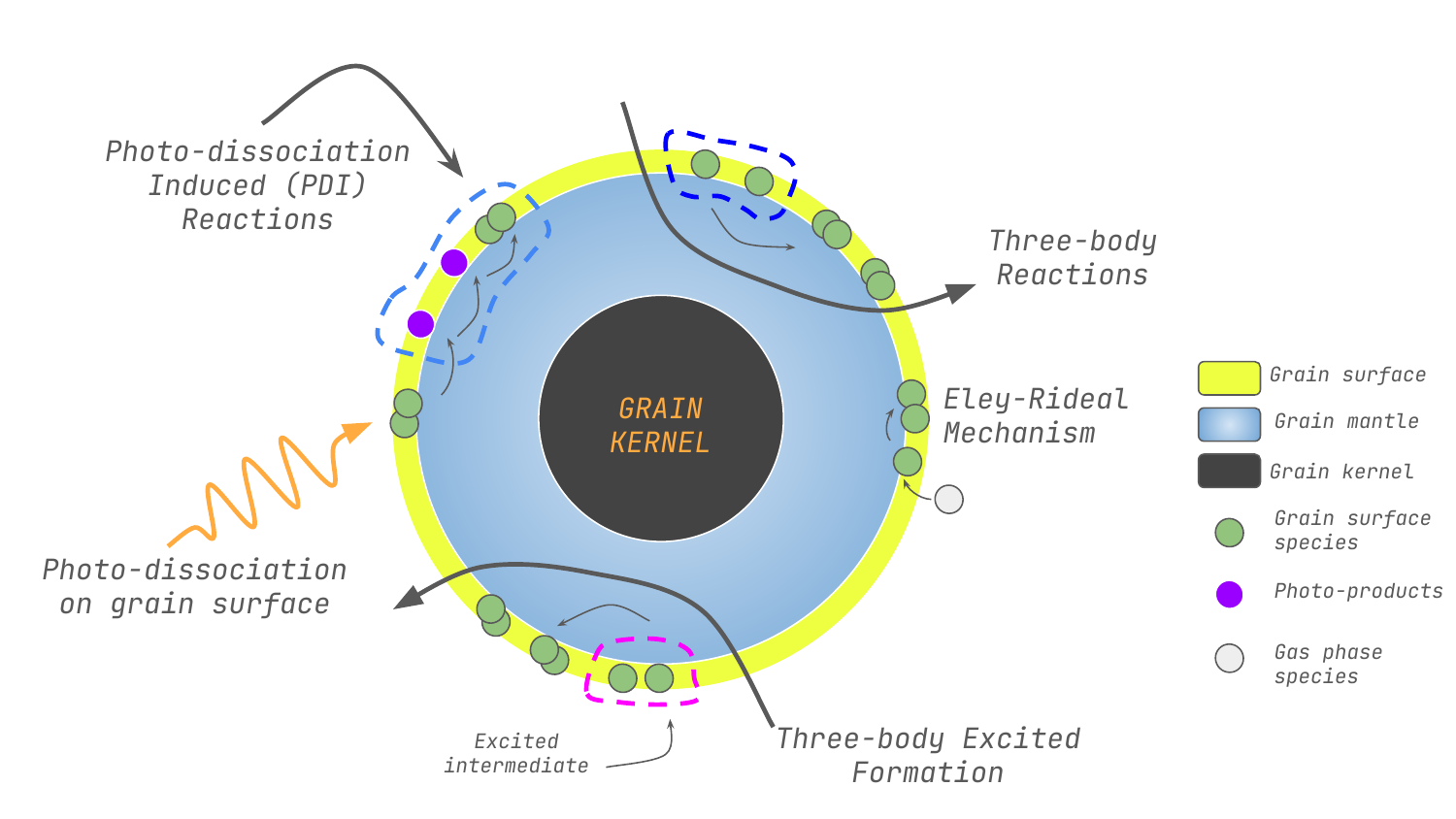}
    \caption{A schematic depicting all non-diffusive chemistry
        processes included in \textsc{Pegasis}.}\label{fig:nondiff}
\end{figure*}

Finally, we implement the non-thermal desorption mechanism introduced by \citet{wakelam2021efficiency}, in which the sputtering of icy grain material due to incoming cosmic ray bombardment is considered. This process allows mantle species to desorb directly into the gas phase without passing through the surface layers.

% All ionization, dissociation, sputtering, and desorption processes are depicted in Figure \ref{fig:dissionizdesorp}.}
% The
% rate formula for this process is given as
% \begin{equation}
%     k_{\text{sp,des}} = \dfrac{\zeta_{\text{CR}}}{3 \times 10^{-17}} \cdot Y_{\text{eff}} \dfrac{\pi r_d^2}{N_s}
% \end{equation}
% with
% \begin{equation}
%     Y_{\text{eff}} = Y^{\infty} \cdot \left[1 - \exp \left\{
%     -{\left(\dfrac{N_{\text{tot}}}{\beta}\right)}^{\gamma}
%     \right\} \right]
% \end{equation}
% Here, \(Y_{\text{eff}}\) is the efficiency of desorption integrated over a
% cosmic ray spectral distribution \citep{dartois2018cosmic}.~\(Y^{\infty}\)
% is the sputtering yield for thick ices with \(\beta\) and \(\gamma\) being
% parameters associated with the nature of ice. We consider sputtering only for
% water ices: \(Y^{\infty} = 3.63; \beta = 3.25; \gamma = 0.57\) \citep{wakelam2021efficiency}.

% \subsection{Diffusive chemistry}\label{sec:diffusive}
\textcolor{red}{Accreted molecules can undergo diffusive reactions via the Langmuir-Hinshelwood (L-H) mechanism. We consider the competition between diffusion, evaporation, and reaction itself, following \citet{Ruaud2016}.} \textcolor{green}{In this mechanism, molecules are \textit{physisorbed}, meaning that weaker van der Waals forces \citep{Reboussin2014} enable diffusion to occur more readily \citep{vidali2006formation}.}
% Once a gas phase molecule has been accreted through process of adsorption onto
% the grain surface, it does not remain stationary. In the case of the
% Langmuir-Hinshelwood (L-H) mechanism, which acts as the primary diffusive
% chemistry driver in our model, two reacting molecules first adsorb onto the the
% grain sites before undergoing a bimolecular reaction on the grain surface. 

% After the process of diffusion, the
% the molecules remain on the surface, with the excess energy being utilized to
% excite the phonon modes of the grain lattice
% \citep{herbst2005chemistry,Semenov2010}.
Physisorbed molecules can diffuse from one site (which is essentially a
potential well) to other in multitude of ways depending on their size. All
molecules can undergo thermally driven diffusion on a time scale given as
\citep{Hasegawa1992,Semenov2010}
\begin{equation}
    t_{\text{hop}} = \dfrac{1}{\nu_{0}} \exp\left(\dfrac{E_b}{T_d}\right)
\end{equation}
where \(T_d\) is the grain temperature.
Following \citet{Hasegawa1992}, we also allow the species to undergo diffusion through quantum tunneling and the time scale \(t_{\text{qt}}\) for this
process is
\begin{equation}\label{eq:tqt}
    t_{\text{qt}} = \dfrac{1}{\nu_0} \exp \left(
    \dfrac{2 a}{\hbar} \sqrt{2 m E_b}
    \right)
\end{equation}
where \(a\) is the diffusion barrier thickness and \(E_b\) is the potential energy barrier between two adjacent surface
sites. Further, we also implement cosmic ray-induced diffusion following \citet{Kalvns2014}.

\subsection{Three-phase Chemistry}

We implement both the prescriptions for three-phase chemistry from
\citet{Ruaud2016} and \citet{hasegawa1993three} which can be switched with each
other as per user's discretion. In \citet{Ruaud2016} (hereafter \citetalias{Ruaud2016}), when a molecule is lost
at the surface, it is immediately replaced by transferring a molecule from the
mantle to the surface, whereas in \citet{hasegawa1993three} (hereafter \citetalias{hasegawa1993three}), the vacant site
remains as is until the determination of the next accretion or desorption
event. If the type of event is accretion, then the vacant site is filled in
with a newly adsorbed molecule, and if it is a desorption event, then the
vacant site is filled in with a molecule from the mantle. More formally, the
net rate of change in total surface material can be written as

\begin{equation}
    \dfrac{d n_{\text{s,tot}}}{dt} = \dfrac{d n_{\text{s,gain}}}{dt}
    + \dfrac{d n_{\text{s,loss}}}{dt}
\end{equation}
Pertaining to this, in \citetalias{hasegawa1993three}, we have
\begin{equation}
    \text{If} \quad \dfrac{dn_{\text{s,tot}}}{dt} > 0
    \begin{cases}
        \dfrac{dn_{\text{s}}(p)}{dt}
        \bigg|_{\text{s} \rightarrow \text{m}} = \alpha_{\text{gain}}
        \dfrac{n_{\text{s}}(p)}{n_{\text{s,tot}}}
        \dfrac{dn_{\text{s,tot}}}{dt} \\ \\
        \dfrac{dn_{\text{m}}(p)}{dt}
        \bigg|_{\text{m} \rightarrow \text{s}} = 0
    \end{cases}
\end{equation}
and
\begin{equation}
    \text{If} \quad \dfrac{dn_{\text{s,tot}}}{dt} < 0
    \begin{cases}
        \dfrac{dn_{\text{s}}(p)}{dt}
        \bigg|_{\text{s} \rightarrow \text{m}} = 0 \\ \\
        \dfrac{dn_{\text{m}}(p)}{dt}
        \bigg|_{\text{m} \rightarrow \text{s}} = \alpha_{\text{loss}}
        \dfrac{n_{\text{m}}(p)}{n_{\text{m,tot}}}
        \dfrac{dn_{\text{s,tot}}}{dt}
    \end{cases}
\end{equation}

Evidently, \citetalias{Ruaud2016} considers accretion at the surface of grains is a
random process; the incoming molecule adsorbs randomly at the surface and the
molecule located below the vacant site is immediately available for desorption.

\subsection{Non-diffusive chemistry}\label{sec:nd}
\begin{figure*}[htbp]
    \centering
    \includegraphics[width=\textwidth]{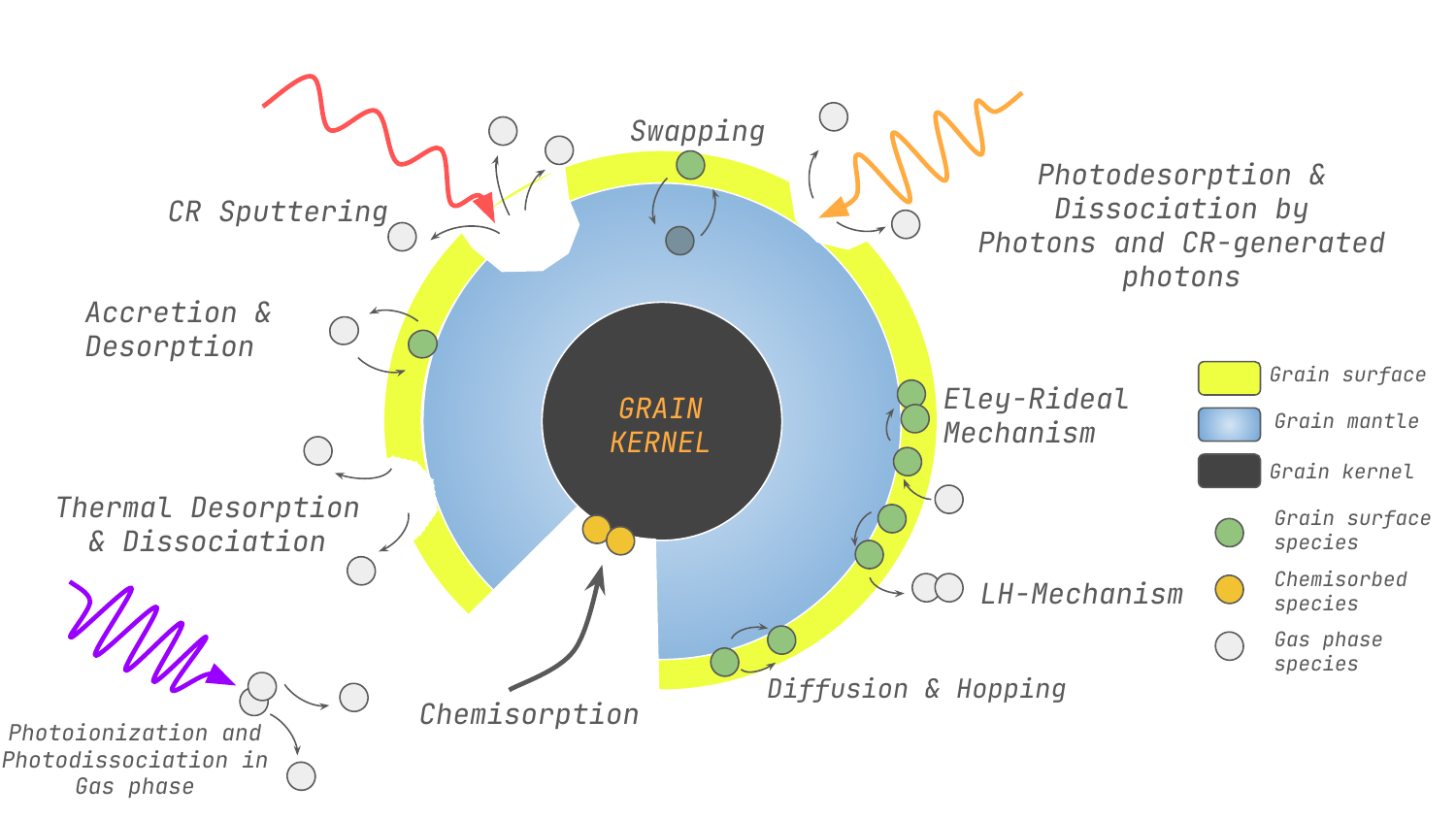}
    \caption{A schematic depicting all ice chemistry
        processes (except non-diffusive chemistry) included in \textsc{Pegasis}.}\label{fig:overview}
\end{figure*}

In order to explain the observations of COMs in cold cores, several authors have proposed alternative formation pathways on grain surfaces without solely relying on diffusive chemistry.
% to produce
% them other than the standard mechanism of recombination of radicals, at
% temperatures around $\qty{30}{\kelvin}$, on ices as the star-forming region heats
% up, eventually being released into the gas phase at temperatures above
% $\qty{100}{\kelvin}$ \citep{Garrod2006}.

% \begin{figure*}[htbp]
%     \centering
%     \includegraphics[scale=0.5]{PEGASIS_2nd_Revision/figures/3bef.png}
%     \caption{Three-body excited formation mechanism that
%         leads to formation of acetaldehyde. We start with the diffusive
%         reaction in step (a), whose exothermicity leads to formation of
%         a vibrationally excited \ce{CH3}. This excited \ce{CH3} then reacts
%         non-diffusively with the readily available \ce{CO} ice after overcoming
%         the activation barrier to form
%         \ce{CH3CO} in step (b). Finally, \ce{CH3CO} gets hydrogenated
%         mainly through the diffusive L-H mechanism to form acetaldehyde in step (c).
%     }\label{fig:3bef}
% \end{figure*}
\citet{ruaud2015modelling} introduced the Eley-Rideal (E-R) mechanism and
formation of the van der Waals complexes on grain surfaces in addition to
standard diffusive chemistry. More recently, \citet{jin2020formation}
formalized several mechanisms further that can proceed through non-diffusive
grain chemistry. In the standard L-H diffusive chemistry formalism of
\citet{Hasegawa1992}, the complete rate equation can be written as
\begin{equation}\label{eq:findiff}
    R^{pq}_{\text{diff}} = \kappa_{pq} (k^{p}_{\text{hop}} + k^{q}_{\text{hop}}) \dfrac{n(p)n(q)}{N_s n_d}
\end{equation}
where \(k_{\text{hop}} = 1/t_{\text{hop}}\) is the hopping rate (thermal or
quantum mechanical tunneling, whichever is faster) with \(n(p)\) and \(n(q)\)
being the abundances of reactants \(p\) and \(q\). In a similar fashion,
we adopt the following general expression for the reactions proceeding
with through the non-diffusive mechanisms \citep{jin2020formation}
\begin{equation}\label{eq:nondiff}
    R^{pq}_{\text{nd}} = \kappa_{pq} R^{p}_{\text{cmp}} \dfrac{n(q)}{N_s n_d} + \kappa_{pq} R^{q}_{\text{cmp}} \dfrac{n(p)}{N_s n_d}
\end{equation}
Here, we have defined \(R^{i}_{\text{cmp}}\) as the \textit{completion} rate of the
reaction which corresponds to the \textit{appearance} rate of species \(i\).
For diffusive chemistry, \(R^{i}_{\text{cmp}} = k^{i}_{\text{hop}} n(i)\) and
equation (\ref{eq:nondiff}) simply reduces to equation (\ref{eq:findiff}). The
expression for \(R^{i}_{\text{cmp}}\) is dependent on the specific
non-diffusive mechanism. All non-diffusive mechanisms are depicted through the schematic in Figure \ref{fig:nondiff} \textcolor{green}{and Appendix \ref{app:gennondiff} describes how the reaction network for non-diffusive reactions was generated from the 2024 KIDA network}.

\subsection{Chemisorption}\label{sec:chemisorption}
% \begin{figure*}[htbp]
%     \centering
%     \includegraphics[width=\textwidth]{PEGASIS_2nd_Revision/figures/ads&diff.png}
%     \caption{A schematic depicting all chemisorption
%         processes in \textsc{Pegasis}.}\label{fig:adsdiff}
% \end{figure*}

So far, in all the chemical processes considered, the surface species were
\textit{physisorbed} from the gas-phase onto the grain surface. As discussed
earlier, molecules are bound to physisorption sites via weak electrostatic
forces which make them susceptible to desorptive processes in high temperature
environments (\(>100\text{ K}\)). To study the effects of surface
chemistry at these temperatures, we include a new variant of surface species in
our model as they behave differently than the species which undergo
physisorption. This different labeling comes as a consequence of treating
physisorption and chemisorption sites differently, with the difference originating
due to the different binding energies with which the molecules bind to the
surface \citep{cazaux2002molecular}. Like non-diffusive chemistry, various
mechanisms through which chemisorption can occur are considered and the
formulation is primarily based on the work of \citet{acharyya2020effect} from
where we also get the reaction set for each of those mechanisms.

Similar to diffusive reactions via physisorption sites, we allow surface
reactions as well as reactive desorption processes through the chemisorbed
sites in mostly an identical manner. Like all mechanisms under chemisorption,
these rates are only calculated when the grain temperature is above
$100\text{ K}$. We implement all the processes as described for physisorbed species so far for each of the diffusive chemistry reaction of the
chemisorbed species included in our network. In summary, all chemical processes (except non-diffusive chemistry) involving grain-surfaces and grain-mantles are depicted in the schematic in Figure \ref{fig:overview}.

\begin{table}
    \caption{Model Descriptions, along with the respective best-fitting times determined using gas-phase observations.}
    \label{tab:model_description}
    \centering
    \resizebox{\columnwidth}{!}{ % Ensures the table fits within a single column
        \begin{tabular}{ccc}
            \toprule
            \textbf{Model} & \textbf{Three-phase Prescription} & \textbf{Non-diffusive Chemistry}  \\
            \midrule
            M1 & \citetalias{Ruaud2016}         & No  \\
            M2 & \citetalias{hasegawa1993three} & No  \\
            M3 & \citetalias{Ruaud2016}         & Yes \\
            M4 & \citetalias{hasegawa1993three} & Yes \\
            \bottomrule
        \end{tabular}
    }
\end{table}

\section{Benchmarking \textsc{PEGASIS}}\label{sec:bench}
% Corresponds to appendix A
We benchmark \textsc{Pegasis} with the public version of \textsc{Nautilus} used in \citetalias{Ruaud2016}. We ran both codes for same physical conditions and switches for the 2014 and 2024 releases of the KIDA network. Note that because many processes such as non-diffusive chemistry and chemisorption are unique to \textsc{Pegasis}, \textcolor{green}{these have been excluded from our comparison models}. To this end, we adopt typical cold core conditions for our physical parameters. \textcolor{red}{The gas number density is taken as \(n_{\ce{H}} = 3\times10^4 \text{ cm}^{-3}\)}. \textcolor{green}{We assume a temperature of $10\text{ K}$ for both gas and grains} and a visual extinction of $15\text{ mag}$. The standard cosmic ionization rate $\zeta_{\ce{H2}}$ of $1.3\times10^{-17}\text{ s}^{-1}$ was assumed. 
% We take the diffusion barrier thickness to be $\qty{1}{\angstrom}$, a typical value for gas-grain models \citep{tielens1982model, Hasegawa1992, Ruaud2016}. 
Additionally, 
we allow photodesorption and self-shielding for \ce{H2}, \ce{CO} and \ce{N2}. Quantum tunneling is not considered \textcolor{green}{in this} comparison; \textcolor{green}{only thermal diffusion being possible}. Spherical grains of radius $0.1 \mathrm{\mu m}$ and density $3\text{ g cm}^{-3}$, with \textcolor{green}{a} dust-to-gas mass ratio of 0.01 were assumed. Following \citetalias{Ruaud2016}, \(10^6\) surface sites \textcolor{green}{per grain} were assumed and \textcolor{green}{the} ratio of diffusion barrier to binding energy is taken \textcolor{green}{to be} 0.4 for surface species and 0.8 for mantle species. Initial elemental abundances are taken from \citetalias{Ruaud2016}. We \textcolor{green}{run} all models for 1 Myr.

We take this opportunity to also use our model to investigate differences between the 2014 and 2024 releases of the KIDA chemical network for key species. \textcolor{green}{The comparison is further discussed in Appendix \ref{app:bench}.} As evident from this benchmarking, \textsc{Pegasis} and \textsc{Nautilus} show excellent agreement for identical initial conditions and chemical networks used to simulate cold core conditions.

% write the parameters and the reason for this choice.
\section{Model Predictions Using Different Approaches}\label{sec:predict}
\begin{figure*}[htbp]
    \centering
    \includegraphics[width=\textwidth]{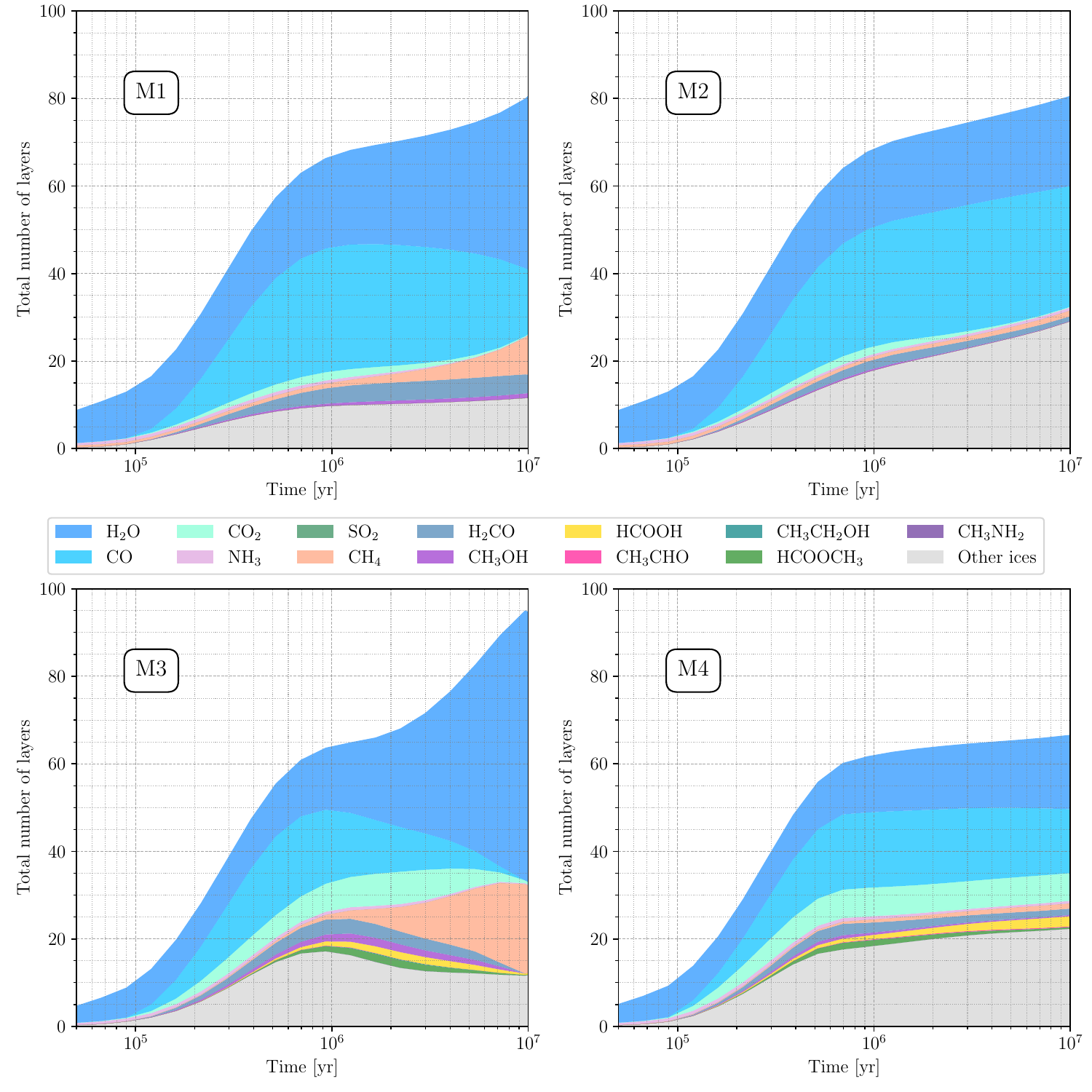}
    \caption{Evolution of ice thickness represented as the total number of layers across major ice species for all the models. The contributions were calculated by accounting for the abundances of both surface and mantle species. The top row is for models with diffusive chemistry and the bottom row for the models with non-diffusive chemistry.}\label{fig:layersplot}
\end{figure*}
To explore the different mechanisms \textcolor{green}{used in the literature to simulate} three-phase chemistry and \textcolor{green}{to consider} the effect of non-diffusive chemistry, we consider four different cold dense cloud models (Table \ref{tab:model_description}). The motivation is to find the model \textcolor{green}{providing} the \textcolor{green}{best} agreement with observed abundances in TMC-1. The \textcolor{green}{physical} conditions \textcolor{green}{assumed are from} \citet{fuente2019gas} \textcolor{green}{with} gas number density \(n_{\ce{H}} = 3\times10^4\text{ cm}^{-3}\), gas and grain temperatures $= 10\text{ K}$, visual extinction $A_V = 15\text{ mag}$ and $\zeta_{\ce{H2}} = 1.3\times10^{-17}\text{ s}^{-1}$. Following \citet{wakelam20242024}, we take a larger value \textcolor{green}{of} $2.5$ \r{A} \textcolor{green}{for the diffusion barrier thickness than was assumed in the benchmark models}. Once again, spherical grains of radius $0.1 \mathrm{\mu m}$ with $3\text{ g cm}^{-3}$ material density are assumed, with a dust-to-gas mass ratio of 0.01. A surface site density of $1.5\times10^{15}\text{ cm}^{-2}$ is assumed. Following \citetalias{Ruaud2016}, the ratio of diffusion barrier to binding energy is taken as 0.4 \textcolor{green}{for the surface and 0.8 for the mantle}. \textcolor{green}{We let all species to diffuse thermally as well as through quantum tunneling. Photodesorption and cosmic-ray sputtering is also enabled for all models. For chemical desorption, we use the prescription from \citet{riedel2023modelling} as it accounts for the nature of the desorbing surface.}  Finally, the initial \textcolor{green}{elemental} abundances are taken in atomic form from \citet{vidal2017reservoir} (except for hydrogen, which is assumed to be converted entirely into molecular form at the beginning) and all models are \textcolor{green}{run} for 10 Myr.

% Discuss importance of initial conditions
\subsection{Comparing Three-phase Chemistry Prescriptions}

\begin{figure*}[htbp]
    \centering
    \includegraphics[width=\textwidth]{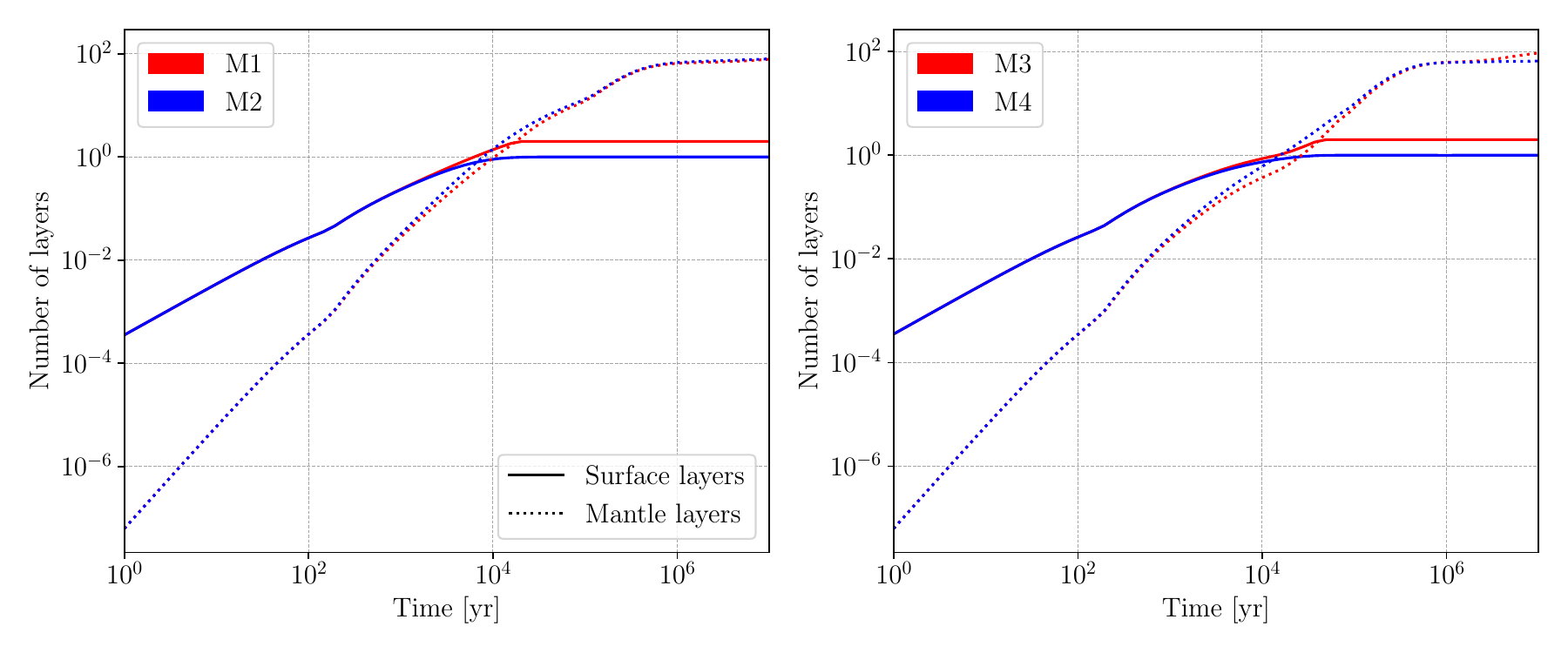}
    \caption{Evolution of ice thickness as a function of number of surface and mantle layers individually for models without non-diffusive chemistry (left) and with non-diffusive chemistry (right).}\label{fig:surfvmantlayers}
\end{figure*}
In this section, we explore \textcolor{green}{the} effect of the two three-phase chemistry prescriptions on \textcolor{green}{the} growth of ices and how non-diffusive chemistry can further influence the contribution of different species. Figure \ref{fig:layersplot} \textcolor{green}{shows} the total number of layers formed by major ices for each model included in this work. \textcolor{red}{The total number of layers is calculated by dividing the ice concentrations (for both surface and mantle) with the total number of sites available per gas phase molecule.} \textcolor{red}{We note that, the original implementation of \citetalias{hasegawa1993three} did not consider active mantle chemistry, but for all our models, mantle species participate in all possible chemical reactions and photoprocesses.}

Both three-phase prescriptions show a very similar distribution of major ices. \ce{H2O} and \ce{CO} ices are the most dominant, making up bulk of the ice thickness. We observed that the diffusive models with \textcolor{green}{both} the three-phase prescriptions of \citetalias{Ruaud2016} and \citetalias{hasegawa1993three} produced similar amount of ice layers, \textcolor{red}{although the share across different ice species is notably different. There is much more \ce{CH4} ice in M1 as compared to M2, and the \ce{CO} abundance does not decline after 1 Myr in case of M2, as it does for M1. Moreover, these trends are similar when non-diffusive chemistry is activated in M3 and M4 (see Section \ref{sec:ndc} for discussion)}. \textcolor{green}{As noted earlier, in the case of \citetalias{Ruaud2016}}, molecules are immediately available for desorption as soon as the molecule above them leaves the surface, independent of the next event. 
% \textcolor{green}{The thermal hopping time-scale of the mantle is what determines the rate of this reaction whereas in the case of \citet{hasegawa1993three}, the mantle concentrations simplyt change as required to balance the material leaving from the surface.}

The individual contributions of major ices are sensitive to different chemical \textcolor{green}{processes that can be included or excluded in a model run and} the chemical network as they affect chemistry by altering dominant pathways of production of these ices. For instance, \citetalias{Ruaud2016}, discussed the effects of reaction diffusion competition on the ice thickness variability. 
% \textcolor{green}{In a similar fashion, we have discussed the role of chemical desorption in case of ice thickness in Appendix \ref{app:icethick}}.

To investigate the differences between the two prescriptions more quantitatively, we looked into the evolution of surface layers and mantle layers individually over the simulation time (Figure \ref{fig:surfvmantlayers}) for models with and without non-diffusive chemistry. \textcolor{red}{We note that the number of mantle layers evolves similarly for both prescriptions, but the surface layers saturate at different values. This is expected, as \citetalias{hasegawa1993three} does not account for surface thickness, whereas \citetalias{Ruaud2016} includes it by considering the two outermost layers as part of the surface, following \citet{fayolle2011laboratory}. In \citetalias{hasegawa1993three}, the active surface layers are modeled through the core coverage factor \(\alpha\)}
\begin{equation}\label{eq:ruaudbeta}
    \alpha_{\text{gain}} = \dfrac{\sum _{p}n_{\text{s}} (p)}{ N_{\text{site}}}
\end{equation}
where the numerator represents the total surface abundance and \(N_{\text{site}}\) is the total number of sites on a grain.
% We note that although the number of mantle layers ends up being roughly equal for both prescriptions as the cold core evolves in time,
% the evolution of number of surface layers highlights the fundamental difference between the two prescriptions. For models M1 and M3, the number of surface layers saturate to a value, which is the assumed number of active layers, here taken as 2 following \citet{fayolle2011laboratory}. This is in agreement with equation (\ref{eq:ruaudbeta}), where it exists to account for surface roughness in \citet{Ruaud2016} but does not in M2 and M4 \citep[see][equation (4)]{hasegawa1993three}. 
% \begin{figure*}[htbp]
%     \centering
%     \includegraphics[width=\textwidth]{PEGASIS_2nd_Revision/figures/2024-diff-nondiff-comparison.pdf}
%     \caption{Same as Figure \ref{fig:surfvmantlayers}, but here models are grouped on the basis of the three-phase prescription used. \citet{Ruaud2016} models M1 and M3 are on the left and \citet{hasegawa1993three} models M2 and M4 are on the right.}\label{fig:diffnondifflayerscomp}
% \end{figure*}
\textcolor{red}{Another key difference between the two prescriptions is that the movement of material between grain surface and grain mantle is solely based on conservation of mass in the case of \citetalias{hasegawa1993three}, whereas \citetalias{Ruaud2016} adopts the recommendation from \citet{garrod2013three}, where these rates are modeled through the thermal hopping (diffusion) time scales of the mantle molecules.} \textcolor{red}{Due to the absence of a buffer where outer monolayers are considered part of the surface in \citetalias{hasegawa1993three}}, the heavier species are progressively trapped in the mantle. At late times, the accretion rates as well as desorption rates decrease as there is not enough material from gas phase to accrete on the grain surfaces. The species locked in mantle have longer desorption time-scales as they have to first transition to the surface. In addition, \textcolor{red}{in case of \citetalias{hasegawa1993three}, the mantle to surface transitions are dependent on the desorption rates, and not the hopping rates, and these rates are much lower at late times when compared to \citetalias{Ruaud2016}}. This affects the heavier species much more than the three lightest species, \ce{H}, \ce{H2} and \ce{He}, which have low binding energies facilitating high desorption rates \citep{govers1980molecular}. \textcolor{red}{In our models, we allow direct desorption from mantles through the process of cosmic-ray sputtering.} This directly brings mantle species back to the gas phase, without them having to first move to the surface layer.
\textcolor{red}{For both prescriptions,} as  adsorption and desorption decline on the surface with time, species still move from the surface to the mantle, thus we observe \textcolor{red}{saturation} of surface abundances after \(t > 10^5\) yr. In the case of \citetalias{Ruaud2016}, the
surface roughness assumption provides a buffer for grain surface chemistry to proceed and saturate after completely
filling up the outer active monolayers region. \textcolor{red}{Finally, we briefly note the the two prescriptions produce significant differences for models with and without non-diffusive chemistry (models M3 and M4 in Figure \ref{fig:layersplot}) which are discussed in more detail in Section \ref{sec:ndc}.} 
\begin{figure*}[htbp]
    \centering
    \includegraphics[width=\textwidth]{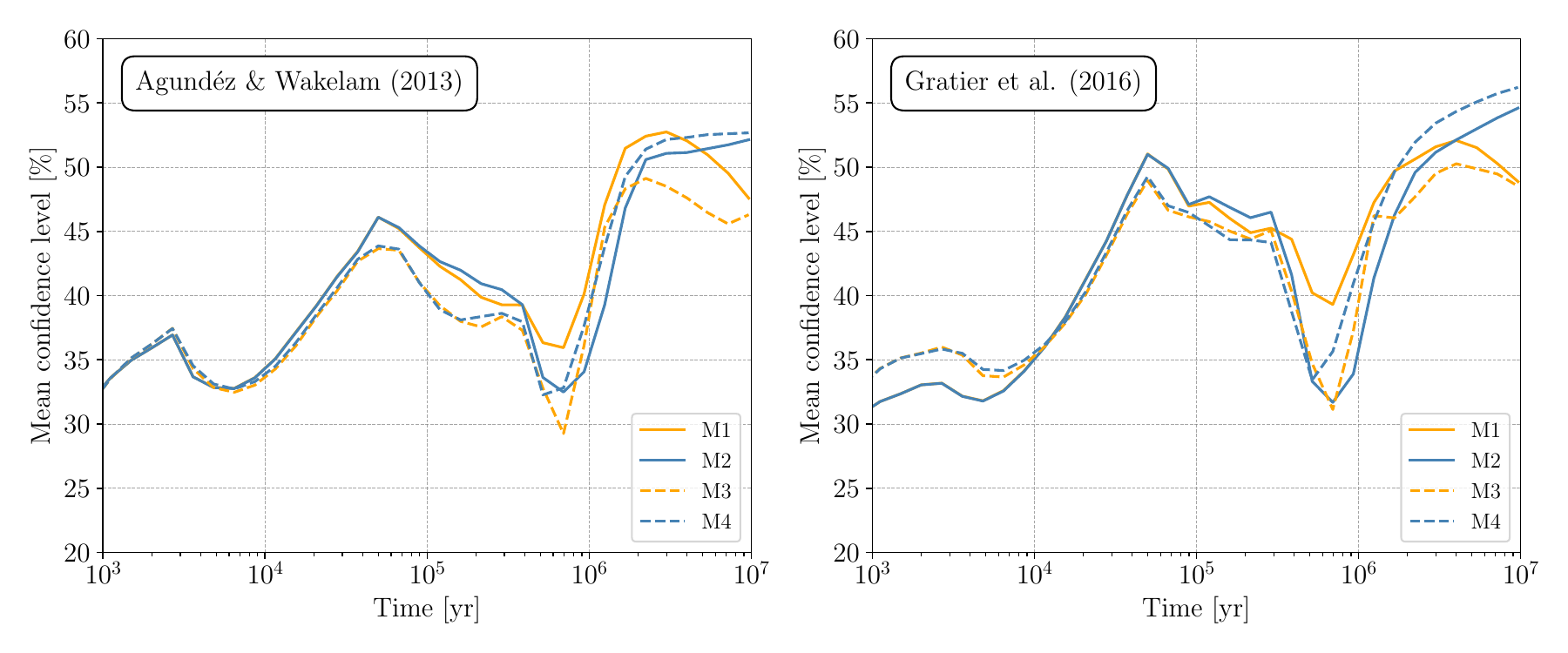}
    \caption{Mean confidence level (\%) for different models for fit with observations from the TMC-1 molecular cloud. The solid lines represent models without non-diffusive chemistry and dashed lines are models with non-diffusive chemistry. 
    % The thicker lines are mean confidence levels with respect to \citet{gratier2016new} observations and the thinner represents the same with respect to \citet{agundez2013chemistry} observations.
    }\label{fig:mcl}
\end{figure*}
\subsection{Gas Compositions in TMC-1}

% \textcolor{red}{Overall, this implies that grain surface-chemistry in the \citet{hasegawa1993three} with an active mantle is mostly similar to \citet{Ruaud2016}, albeit with lower overall surface concentrations}. 

%According to experimental work by \citet{watanabe2007laboratory} and \citet{ioppolo2010water}, two-body association reactions which lead to form ices are more likely to occur on the grain-surface rather than inside the bulk mantle. Note that both of these prescriptions assume a level of homogeneity in the surface and mantle components, which is to say that the monolayers are not treated differentially according to their depth.~\citet{furuya2017water} adopted a slightly different approach in which a depth-dependent seven-phase model was designed as an extension of \citet{hasegawa1993three}. \textcolor{green}{This model introduced partitioning of mantle layers into groups called phases to simulate inhomogeneity.} 

% Comparing HH93 and Ruaud
% Layer figures

% From Ruaud slides

% \begin{figure*}[htbp]
%     \centering
%     \includegraphics[scale=0.7]{PEGASIS_2nd_Revision/figures/abun.pdf}
%     \caption{Time evolution of acetalydehyde in gas phase
%         and as ice using \textsc{Pegasis}.}\label{fig:acetaldehyde}
% \end{figure*}

In this section, we concentrate on reproducing the gas-phase chemical abundances \textcolor{green}{observed in} the well studied TMC-1 (CP) \citep{kaifu20048, gratier2016new, agundez2013chemistry}. All observations were derived in the form of column densities and \textcolor{green}{then} converted to abundances with respect to atomic hydrogen assuming \(N(\ce{H2}) = 10^{22}\text{ cm}^{-2}\) \citep{cernicharo1987physical, gratier2016new}.

\textcolor{green}{The models in Table \ref{tab:model_description} share the same physical and chemical conditions but differ in their choice of three-phase chemistry prescription and the inclusion or exclusion of non-diffusive chemistry. The goal is to determine which of the four models best reproduces the observations and to estimate the chemical age of the source \citep[see, for instance,][]{Majumdar17}. According to \citet{smith2004rapid} and \citet{wakelam20242024}, abundances are considered well-reproduced if they fall within one order of magnitude of the observed value. There have been several statistical methods employed by various authors to get an estimation on how well an observation is reproduced by a model. \citet{wakelam20242024,wakelam2010reaction} employ the distance of disagreement method, where the best-fitting time corresponds to the minimum value of the parameter \(D\).
}
\begin{equation}
    D(t) = \dfrac{1}{N_i} \sum_i \left | \log(X_{\text{mod},i}(t)) - \log(X_{\text{obs},i}) \right |
\end{equation}
where \(X_{\text{mod},i}(t)\) is the modeled abundance of species \(i\) at time \(t\) and \(X_{\text{obs},i}\) is the observed abundance of species \(i\) from an observation dataset (\citet{agundez2013chemistry} or \citet{gratier2016new}). This works well to get an estimate on the best-fitting time but is susceptible to (large) outliers and would be better suited if the uncertainties on all observations were readily available. The mean logarithmic differences approach from \citet{loison2013gas} and \citet{wakelam20152014} suffers from a similar problem.

\textcolor{green}{Instead}, we use the mean confidence level approach discussed in \citet{garrod2007non}, which works well for our case and has also been applied in \citet{Majumdar17} and \citetalias{Ruaud2016}. For this, we define \textit{confidence level} \(\kappa_i\) as
\begin{equation}
    \kappa_i = \text{erfc} \left( \dfrac{\log(X_{\text{mod},i}) - \log(X_{\text{obs},i})}{\sqrt{2}\sigma} \right)
\end{equation}
where \(\text{erfc}\) is the complementary error function and we take \(\sigma = 1\) implying 1 standard deviation corresponds to the modeled value being within one order of the observed value following \citet{garrod2007non}. We calculate \(\kappa_i\) for each species at each time and \textcolor{green}{the} \textit{overall} confidence at each time is defined by taking \textcolor{green}{the} mean of the individual levels over all species. We plot the mean confidence level for each model in Figure \ref{fig:mcl} for both the observation datasets. 
% \textcolor{red}{LM: WHY? For this analysis, we (re)-simulated all models in Table \ref{tab:models} for a total of 100 Myr, ensuring all models had reached a steady state.}
For this analysis, we have \textcolor{green}{considered} all \textcolor{green}{the} models in Table \ref{tab:model_description} over a period of 10 Myr.
% , although we note that not all models yield a definitive best-fitting time by then. Following a similar approach to that of \citet{garrod2007non}, we deem such very late times highly suspect. 
% \textcolor{red}{Still, we note that model M2 does not reach a steady state within the simulation time.}

\begin{figure*}[htbp]
    \centering
    \includegraphics[width=\textwidth]{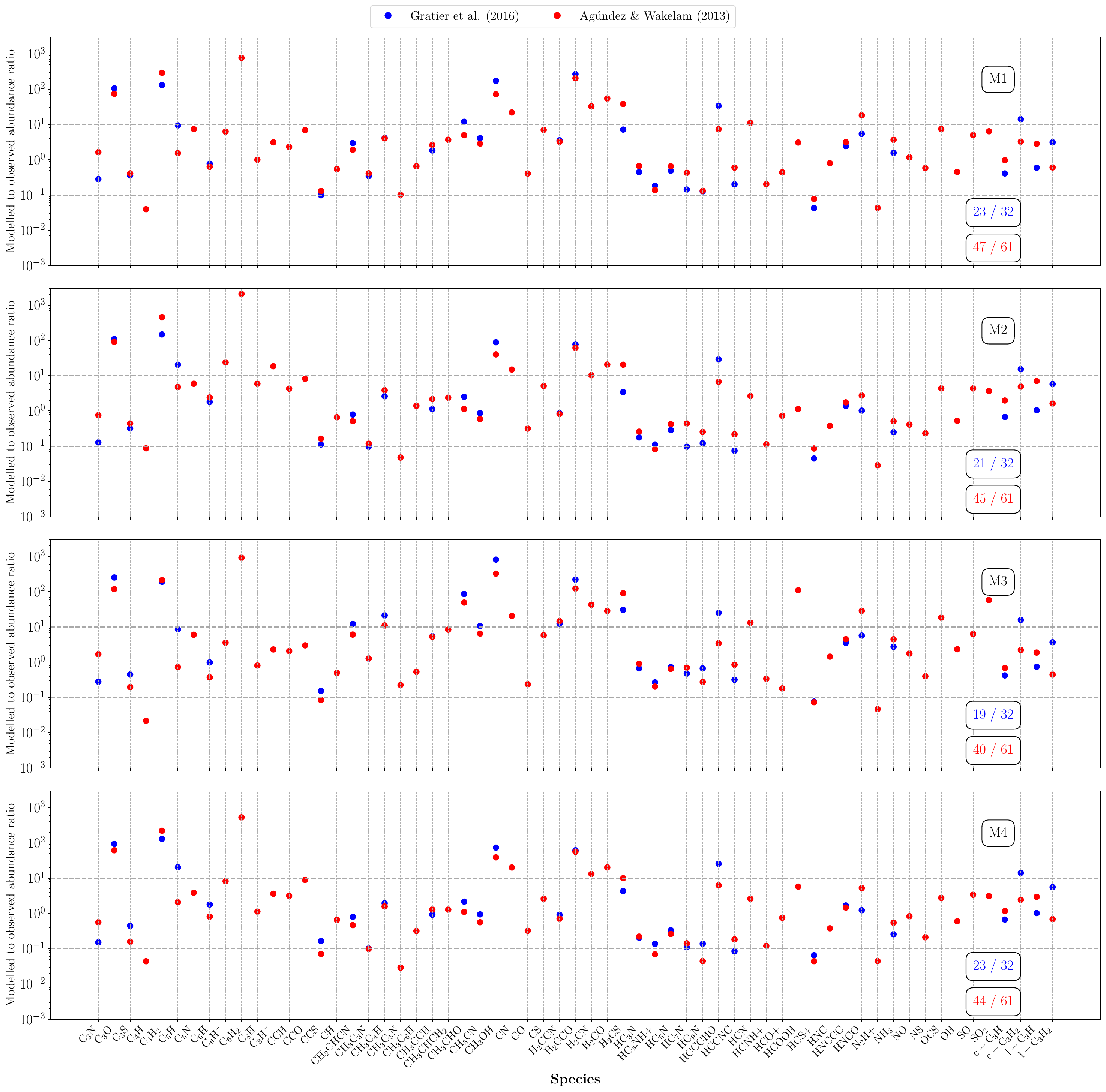}
    \caption{The ratios of modeled abundances to observed abundances for both datasets used in this work for each model. Only species without any upper or lower limits are displayed here, with the number of species that each model could reproduce for the datasets of \citet{gratier2016new} (in blue) and \citet{agundez2013chemistry} (in red).}  \label{fig:dexwithin}
\end{figure*}

Our results \textcolor{green}{are similar to} the findings of \citet{garrod2007non} in that they produce two distinct local maximas for the mean confidence level, and therefore provide two choices for the best-fitting time. The first local maximum for all models with respect to both observation sets is much flatter and represents a considerably larger interval ranging from approximately \(4 \times 10^4\) yr to \(2 \times 10^5\) yr. On the other hand, the late times \textcolor{green}{obtained for models M1 and M3 (few \(10^6\) yr)} produce a greater mean confidence level value, though \citet{hartquist2001chemical} suggest that TMC-1 is a relatively young molecular cloud \textcolor{green}{(\(< 10^5\) yr for TMC-1 Core D)}. However, according to \citet{hartmann2001rapid} and \citet{mouschovias2006observational}, an age of a few \(10^5\) yr is too short as some estimates put the age of these dense clouds in the range of \(10^6-10^7\) yr. In a more recent work, \citet{navarro2021evolutionary} suggest that observations of TMC-1 (CP) can be best explained with 
% \textcolor{red}{LM: I THINK IT IS 1 MYR:  \(t\sim10\) Myr} 
a gravitational collapse model (\(t\sim1\) Myr) \textcolor{green}{or} a more sophisticated collapse model with ambipolar diffusion (\(t\sim10\) Myr). Nevertheless, we note that the late best-fitting times for models M1 and M3 are in agreement with the best times found by \citet{wakelam20242024}. 
\textcolor{red}{For the \citetalias{hasegawa1993three} (M2 and M4) models we fail to find a definitive maximum at late times with either of the observational datasets of \citet{agundez2013chemistry} or \citet{gratier2016new}. 
% is not reached when using the observations from \citet{gratier2016new}, and occurs very late (\(\sim 4 \times 10^7\) yr) when using observations from \citet{agundez2013chemistry}. 
For M1 and M3 (i.e., the \citetalias{Ruaud2016} models), where we do observe a peak after 1 Myr, we note that the confidence level value for these late times exceed the value for early time.} 

% \textcolor{red}{Recent works like \citet{vazquez2018molecular} and \citet{jeffreson2021scaling} have attempted to estimate molecular cloud lifetimes and find that the typical timescales lie within 100 Myr or so. Therefore, it is essential for the gas-grain models to allow all the species to attain a steady state within that time-frame. 
The cases where we note the confidence still increasing beyond 10 Myr suggests at the limitations of our model. These limitations can be attributed to both the chemical network, as well as the individual chemical processes considered in our model. Moreover, our model is a pseudo-time-dependent, where the physical conditions remain constant through the evolution of the cloud. In reality, these environments are much more complex, with processes like turbulence, shocks and evolving densities due to core collapse affecting the chemical time-scales immensely. Therefore, in the cases where the statistical methods fail to produce the best time within the reasonable limits, authors such as \citet{garrod2007non} have opted to discard the models on physical and chemical grounds. The effects of non-diffusive chemistry in estimating chemical ages within the 10 Myr range is discussed in Section \ref{sec:ndc}.

\begin{table*}
    \caption{Observed ice composition in various massive young stellar objects (MYSOs), low-mass young stellar objects (LYSOs), and background (BG) stars, along with a Kuiper belt comet, compared against model predictions at the best-fitting times determined using gas-phase abundances.}\label{tab:species_data}
    \centering
    \small
    \resizebox{\textwidth}{!}{ % Ensures the table fits within text width
    \begin{tabular}{lcccccccc}
        \toprule
        \multirow{2}{*}{\textbf{Ice Species}} & \multicolumn{4}{c}{$X_{\ce{H2O}} (\%)$} & \multicolumn{4}{c}{Literature ($\%$ \ce{H2O})} \\
        \cmidrule(lr){2-5} \cmidrule(lr){6-9}
        & M1 & M2 & M3 & M4 & BG Stars & MYSOs & LYSOs & Comets \\
        \midrule
        \ce{H2O\tablefootmark{$\star$}}     & $9.28(-5)$  & $6.95(-5)$    & $1.16(-4)$   & $5.71(-5)$  & 100      & 100         & 100          & 100    \\
        \midrule
        \ce{CO}      & 91.84  & 134.92    & 18.68   & 87.17  & 9--67\tablefootmark{(a)}    & 3--26\tablefootmark{(a)}     & $<$3--85\tablefootmark{(a)}     & 0.4--30\tablefootmark{(a,e)}    \\
        \ce{CO2}     & 2.84    & 0.76     & 16.96   & 37.63   & 14--43\tablefootmark{(a)}   & 11--27\tablefootmark{(a)}      & 12--50\tablefootmark{(a)}       & 4--30\tablefootmark{(a,e)}      \\
        \ce{CH4}     & 12.70    & 6.36     & 32.53   & 7.12    & $<$3\tablefootmark{(a)}     & 1--3\tablefootmark{(a)}        & 1--11\tablefootmark{(a)}        & 0.4--1.6\tablefootmark{(a,e)}   \\
        \ce{NH3}     & 0.86    & 3.37     & 1.38    & 3.35    & $<$7\tablefootmark{(a)}     & $\sim$7\tablefootmark{(a)}     & 3--10\tablefootmark{(a)}        & 0.2--1.4\tablefootmark{(a,e)}   \\
        \ce{CH3OH}   & 2.93    & 0.45     & 3.98    & 1.59    & $<$1--12\tablefootmark{(a)} & $<$3--31\tablefootmark{(a)}    & $<$1--25\tablefootmark{(a)}     & 0.2--7\tablefootmark{(a,e)}      \\
        \ce{H2CO}    & 15.88    & 5.31     & 7.04   & 8.64    & --       & $\sim$2--7\tablefootmark{(a)}  & $\sim$6\tablefootmark{(a)}      & 0.11--1.0\tablefootmark{(a,e)}       \\
        
        \ce{HCOOH}   & $5(-3)$ & $3(-3)$  & 4.24    & 13.69   & $<$2\tablefootmark{(a)}     & $<$0.5--6\tablefootmark{(a)}   & $<$0.5--4\tablefootmark{(a)}    & 0.06--0.14\tablefootmark{(a,e)}      \\
        \ce{CH3CHO}  & $1(-3)$ & $5(-5)$  & 0.02    & 1.03    & --       & $<$2.3\tablefootmark{(b)}       & --           & 0.047\tablefootmark{(f,g)}      \\
        \ce{C2H5OH}  & $4(-4)$ & $2(-4)$  & 0.31    & 0.13    & --       & $<$1.9\tablefootmark{(b)}      & --           & 0.039\tablefootmark{(f,g)}      \\
        \ce{HCOOCH3} & $0.06$    & $0.02$  & 2.98    & 2.14    & --       & --          & $<$2.3\tablefootmark{(c)}       & 0.0034\tablefootmark{(f,g)}     \\
        \ce{CH3NH2}  & $0.01$    & $1(-3)$  & 0.08    & 0.19    & --       & $<$3.4\tablefootmark{(b)}      & $<$16\tablefootmark{(d)}        & --         \\
        % \ce{H2O2}    & 9.25    & 1.48     & 12.41   & 9.62    & --       & $\sim$2--7  & $\sim$6      & 0.32       \\
        % \ce{OCS}    & 9.25    & 1.48     & 12.41   & 9.62    & --       & $\sim$2--7  & $\sim$6      & 0.32       \\
        \ce{SO2}     & $3(-5)$ & $6(-10)$ & $6(-4)$ & $5(-3)$ & --       & $<$0.9--1.4\tablefootmark{(a)} & 0.08--0.76\tablefootmark{(a)}   & 0.2\tablefootmark{(a,e)}        \\
        \bottomrule
    \end{tabular}
    }
    \tablefoot{Model columns (M1--M4) represent specific simulation setups where \(a(b) = a \times 10^b\), and the observation columns summarize results from different astronomical sources. Dash (--) indicates no data. The best-fitting times for models M1-M4 were 4 Myr, 9.64 Myr, 4 Myr and 9.64 Myr respectively.
    \\
    \tablefoottext{$\star$}{For \ce{H2O}, we have listed the abundance of total water ice with respect to gas-phase \ce{H} for each model.}
    }
    % c-> b
    % d -> c
    % g -> d
    % f -> e
    % b -> f
    % e -> g
    \tablebib{
    \tablefoottext{a}{\citet{boogert2015observations},}
    \tablefoottext{b}{\citet{terwisscha2018infrared},}
    \tablefoottext{c}{\citet{van2021infrared},}
    \tablefoottext{d}{\citet{rachid2020infrared},}
    \tablefoottext{e}{\citet{mumma2011chemical},}
    \tablefoottext{f}{\citet{rocha2024jwst},}
    \tablefoottext{g}{\citet{rubin2019elemental}.}
    }
\end{table*}

Overall, \textcolor{red}{as evident from Figure \ref{fig:dexwithin}}, the \citetalias{Ruaud2016} model M1, can reproduce the observed abundances for the most number of species, 23 out of 32 gas-phase species abundances for \citet{gratier2016new} and 47 out of 61 gas-phase species from \citet{agundez2013chemistry} \textcolor{red}{with best times of 4 Myr and 3 Myr respectively} within a factor of ten. Note that while making Figure \ref{fig:dexwithin}, any observations with lower or upper limits were not considered. Upon including them, M1 can reproduce 28 out of 37 gas-phase species abundances for \citet{gratier2016new} (with a confidence of 52\%) and 48 out of 70 gas-phase species from \citet{agundez2013chemistry} (with a confidence of 53\%). The non-diffusive \citetalias{Ruaud2016} model M3 also produces similar best-fitting times as M1, although with less mean confidence. \citetalias{hasegawa1993three} models M2 and M4, despite producing even larger confidence (as high as \(56\%\) in case of M4) at late times, failed to attain a maximum within 10 Myr.

% \textcolor{red}{BE SPECIFIC, WHICH MODEL, HOW MANY, ALSO ACCURATELY OR WITHIN A FACTOR OF 10, WHICH CONSIDERED AS GOOD FIT IN ASTROCHEMISTRY. }

This ambiguity \textcolor{red}{while trying to find the best-times} is expected because of the complex structure of TMC-1 as well as \textcolor{green}{the} distinct physical history of its substructures. The time where the mean confidence attains a maximum is also susceptible to the initial conditions assumed, especially the starting \ce{C}/\ce{O} ratio \citep{agundez2013chemistry}. The role of desorption mechanisms in estimating the chemical ages through modeling has been brought up several times \citep{herbst1995chemistry,Herbst2001,garrod2006gas}.
Finally, the networks can also affect the overall agreements with the observations and provide different best-fitting times \citep{wakelam20242024}. 
% \textcolor{red}{We explore the effects of chemical desorption mechanism
% and the reaction network in Appendix \ref{app:age}.}

% We also did a comparison with the data reported in \citet{agundez2013chemistry} which can be found in the Appendix. Here
% \begin{figure*}[htbp]
%     \centering
%     \includegraphics[scale=0.35]{PEGASIS_2nd_Revision/figures/dod.pdf}
%     \caption{Ratio of modelled abundances to the observed abundances at
%     the best fitting-time with M4, the model with smallest distance of disagreement.}\label{fig:dod}
% \end{figure*}

\subsection{Ice Compositions}
\begin{figure*}[htbp]
    \centering
    \includegraphics[width=\textwidth]{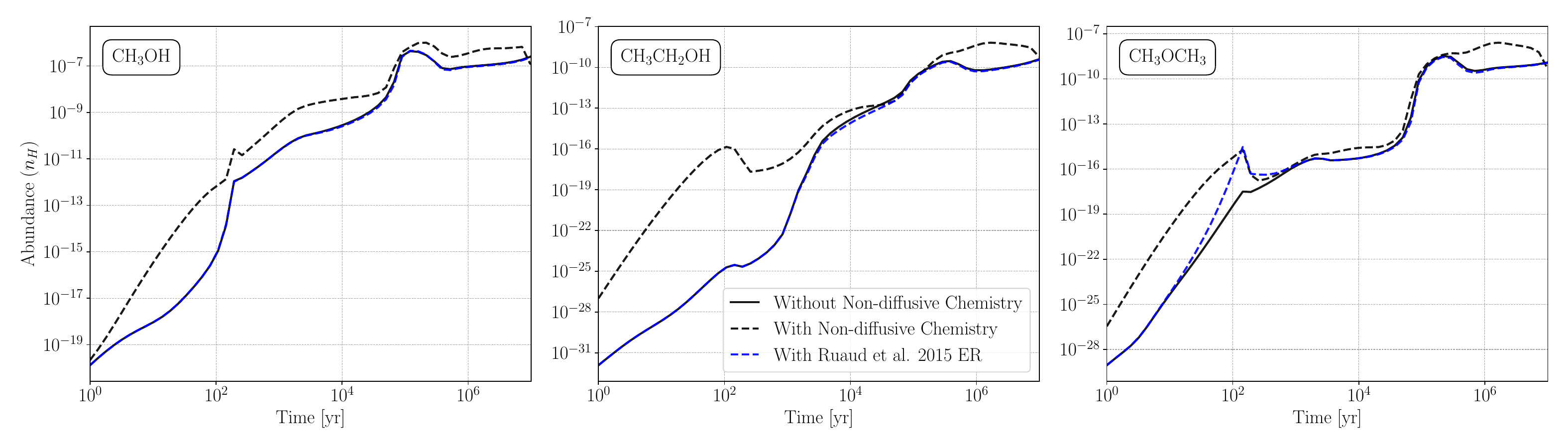}\\
    \includegraphics[width=\textwidth]{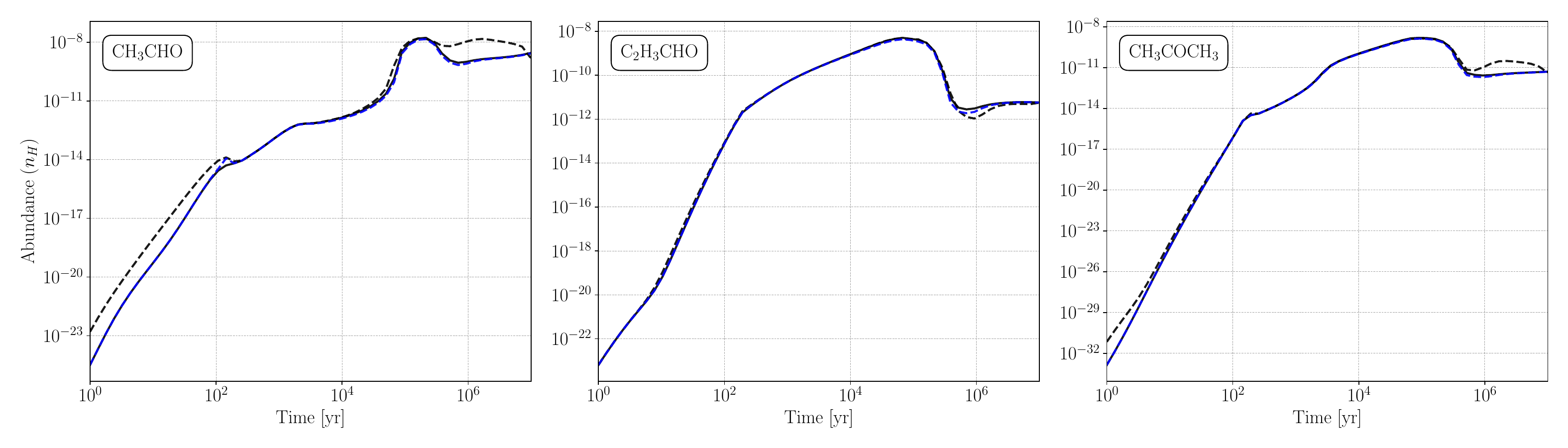}\\
    \includegraphics[width=\textwidth]{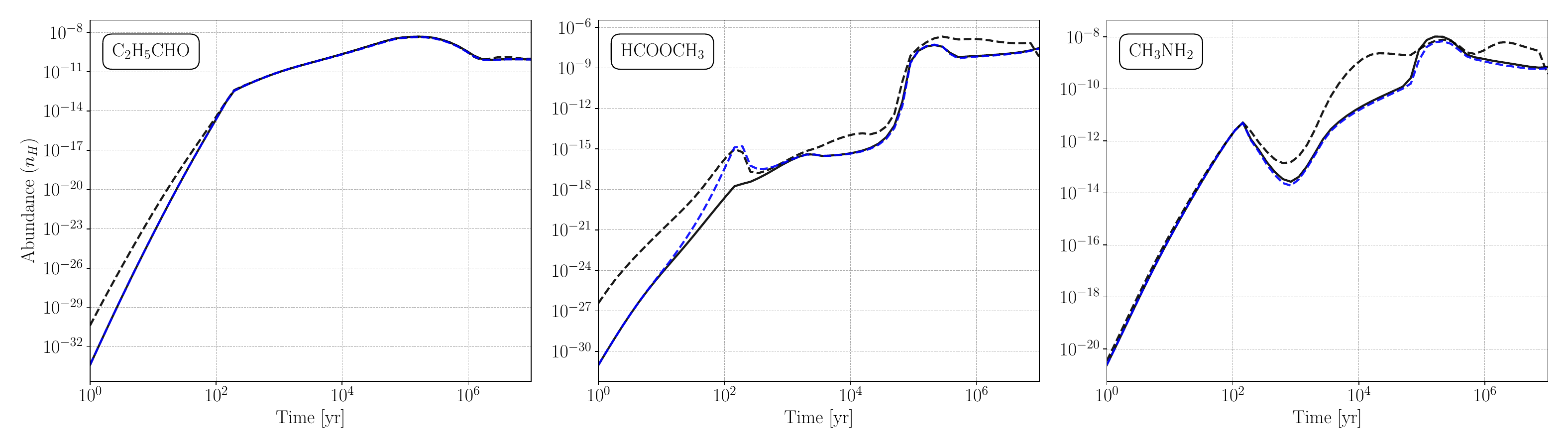}
    \caption{Abundances of major COMs in gas-phase as a function of time with (model M3) and without (model M1) the inclusion of non-diffusive chemistry.}\label{fig:majorcomsevolgas}
\end{figure*}

Ice compositions have been observed in absorption along the line of sight towards background stars, massive young stellar objects (MYSOs) and low-mass young stellar objects (LYSOs). Here we compare observations of these objects with our model predictions. We use the ice abundances at the best fit age for gas phase observations of TMC-1. The observations and model abundances (relative to those of water ice) are given in Table \ref{tab:species_data}.  We include both surface and mantle abundances in the model predictions.  In addition to the interstellar sources, we also include the abundances from cometary bodies in our comparison.

%We also compare the abundances of prominent ices and COMs against observations in slightly more evolved sources and objects than molecular clouds such as background stars and young stellar objects as well as the comet 67P/C-G (from the Kuiper belt). The results are tabulated in Table \ref{tab:species_data}. We use the best-fitting times found using the gas phase abundances to obtain the corresponding ice abundances. \textcolor{red}{The best-fitting times for all models are tabulated in Table \ref{tab:model_description}.} We consider contribution from both surface and mantle portions that make up the total ice abundance. 

In this section, we discuss the results of model M1 \textcolor{red}{and M2} for major ices species. Models M3 and M4 which involve non-diffusive chemistry are discussed in the next section. \textcolor{red}{Table \ref{tab:species_data} clearly shows that non-diffusive chemistry can significantly impact the contribution of different ice species to the total ice, the differences are evident between M1 and M3 and between M2 and M4}. The sources and objects \textcolor{green}{considered in this section} are more evolved than TMC-1 and have different physical conditions.  They have also been the subject to significant processing and evolution as those physical conditions change, hence we do not expect \textcolor{green}{good} agreement \textcolor{green}{with the models}. However, this comparison highlights how different the ice inventories are for molecular clouds when compared to these sources. This can give clues \textcolor{green}{as to} how different species are being preserved, destroyed or converted into other species as the cloud evolves and collapses to form proto-stellar objects.   

In model M1, we observe \ce{CO} ice is almost at par with \ce{H2O}, which is expected in certain scenarios in dense clouds. This is in agreement with \citetalias{Ruaud2016}, for which \ce{CO} ice was produced at par with \ce{H2O} ice when competition between diffusive reaction, desorption and accretion was taken into account. In fact, in M2, the abundance of \ce{CO} ice far exceeds \ce{H2O}. In our case, as a result of assuming the diffusion barrier thickness as $2.5$ \r{A} \citep{wakelam20242024}, the reaction path of \ce{OH + CO -> CO2 + H} is more efficient by \textcolor{red}{two} orders than the reaction \ce{O + HCO -> CO2 + H} in agreement with \citet{garrod2011formation}. However, this is still not enough to drive the conversion of \ce{CO} ice to \ce{CO2} as accretion of \ce{CO} is two orders of magnitude faster than accretion of \ce{CO2} at this time. Note that for all of the ice species in model M1, the \textcolor{green}{fastest} rate is almost always for the movement of molecules from the ice surface to the mantle and vice versa. The over-prediction of \ce{CH4} ice is also a matter of interest. \ce{H + CH3 -> CH4} emerges as the most \textcolor{green}{important} reaction forming \ce{CH4} ice on the grain surface which is also the major route of formation as described in the experimental work of \citet{qasim2020experimental}. In contrast, model M2 does not exhibit this over-prediction of ice. Upon further inspection, this can be attributed to inefficient movement of the ices from mantle to the surface at late times in the \citetalias{hasegawa1993three} model. The disparity between the abundance of \ce{H2CO} and other ices in Table \ref{tab:species_data} between models M1 and M2 can be explained likewise.

The differences in abundances detected in the sources in Table \ref{tab:species_data} suggest that at the best-fitting time, this model had not attained the expected ratios for most ices with respect to \ce{H2O} ice. Further, note that these ratios only give half of the picture. The abundances relative to \(n_{\ce{H}}\), can also be derived from the corresponding column densities, as  in \citet{boogert2015observations}. Our model predicts these abundances for TMC-1 within one order of magnitude \textcolor{green}{of those observed}, suggesting that over the course of the evolution from clouds to proto-stellar objects, the changes occur mostly in the ratios in which these species exist as ices.

\subsection{Effects of Non-Diffusive Chemistry}\label{sec:ndc}
% Comparison with Ruaud's ER
% Discussion on Reactive desorption
% More discussion on late times, link to desorptive processes+analysis through elemental ratios (see agundez+wakelam)
% \begin{figure*}[htbp]
%     \centering
%     \includegraphics[width=\textwidth]{PEGASIS_2nd_Revision/figures/['CH3OH', 'HCOOH']-ices-abun.pdf}
%     \caption{Abundances of \ce{CH3OH} and \ce{HCOOH} ice as a function of time with and without the inclusion of non-diffusive chemistry.}\label{fig:ch3ohhccoh}
% \end{figure*}
% \begin{figure*}[htbp]
%     \centering
%     \includegraphics[width=\textwidth]{PEGASIS_2nd_Revision/figures/['CH3CHO', 'CH3OCH3']-ices-abun.pdf}
%     \caption{Abundances of \ce{CH3CHO} and \ce{CH3OCH3} ice as a function of time with and without the inclusion of non-diffusive chemistry.}\label{fig:ch3chochoch3}
% \end{figure*}

Non-diffusive chemistry facilitates the formation of COMs. As evident from Figure \ref{fig:layersplot}, it does not necessarily increase the amount of ices, rather, it assists in the formation of COMs from precursors like methanol. Ices like \ce{CH4}, \ce{CH3OH}, \ce{HCOOH} and \ce{HCOOCH3} are boosted for both models M3 and M4  (Figure \ref{fig:layersplot}). \textcolor{red}{In fact, we see that the number of total ice layers decreases in the case of the \citetalias{hasegawa1993three} three-phase chemistry prescription and increases in the case of \citetalias{Ruaud2016}. For M3,  water ice sees a rapid build-up shortly after 1 Myr of evolution, something which is absent in model M4. Specifically, the non-diffusive photodissociation-induced reactions in M3 are more efficient in producing water ice than in M4 in the grain mantle. The formation of water ice can be attributed to the following reactions}
\vspace{-2em}
\begin{center}
    \[
        \ce{b-H2 + b-OH -> b-H2O + b-H} \tag{R1} \label{reac:effnondiffwater}
    \]
    \[
        \ce{b-CH4 + b-OH -> b-CH3 + b-H2O} \tag{R2} \label{reac:effnondiffmethane}
    \]
\end{center}
The prefix `b-' denotes a species present in the bulk (mantle). Reaction~\ref{reac:effnondiffwater} follows a diffusive pathway, as \ce{b-H2} is light enough to diffuse readily in cold environments (approximately $\sim 10\text{ K}$). In contrast, Reaction~\ref{reac:effnondiffmethane} primarily proceeds via non-diffusive pathways. Both of these reactions exhibit significantly lower rates in model M4. The lack of enough methane ice in the \citetalias{hasegawa1993three} models is one of the major reasons for which we observe this as Reaction \ref{reac:effnondiffmethane} is not as efficient. Further, the inclusion of non-diffusive chemistry boosts abundance of \ce{CO2} ice for both the models M3 and M4 as its production no longer relies on the very slowly diffusing \ce{CO} and \ce{O} at such low temperature of $10\text{ K}$. Non-diffusive chemistry enables formation of \ce{CO2} without the requirement of diffusion, in agreement with the findings of \citet{jimenez2025modelling} where models with non-diffusive chemistry could produce of \ce{CO2} ice even at low temperatures of \(T_{\text{dust}} < 12\, \mathrm{K}\).

Figure \ref{fig:surfvmantlayers} shows that non-diffusive chemistry also does not affect the split between surface and mantle abundances given a three-phase prescription. \textcolor{red}{Enabling non-diffusive chemistry not only makes significant differences in the abundances of major ice species across all models (see Table \ref{tab:species_data}), it also boosts the abundances of COMs in each of them.  Non-diffusive chemistry in the case of \citetalias{hasegawa1993three}, could lead to a best-time closer to previous estimates of few Myr. This may be attributed to the fact that non-diffusive chemistry provides extra pathways forming COMs, which are heavier and are susceptible to getting trapped in the mantle, nullify the lesser available concentrations of surface species in the case of \citetalias{hasegawa1993three}. Noting that the effects and trends of non-diffusive chemistry are similar across all models, we limit our discussion and results to the \citetalias{Ruaud2016} models for brevity.} 
% \textbf{Further, in order to investigate the influence of non-diffusive mechanisms, we examine evolution of abundances of COMs that have been detected in gaseous and ice phase in cold cores, with emphasis on the object of our interest--TMC-1.}

\subsubsection{On COMs in the gas-phase}
% As mentioned earlier, several COMs have been detected in TMC-1 but we limit our discussion to species which have been detected in the
Figure \ref{fig:majorcomsevolgas} shows abundances of various COMs in gas-phase that have been detected in TMC-1 and are included in the 2024 KIDA network. This evolution of abundances can be thought to occur over three epochs: early stage (from beginning till $10^2$ yr), the middle stage (from $10^2$ yr to $10^4$ yr) and finally the late stage (beyond $10^4$ yr). Owing to three distinct phases of evolution, we note there are non-trivial effects on abundances of methanol (\ce{CH3OH}), ethanol (\ce{CH3CH2OH}), dimethyl ether (\ce{CH3OCH3}) and methyl formate (\ce{HCOOCH3}) in the early stage, with rest of species having close to marginal differences between the models with and without non-diffusive chemistry. In the middle stage, ethanol exhibits the greatest deviation from the diffusive-only chemistry once non-diffusive mechanisms are included. In the late stage, which also contains the best-fitting time, the differences for all species gradually decrease as all abundances converge to steady-state. We note that for species like propenal (\ce{C2H3CHO}), acetone (\ce{CH3COCH3}) and propanal (\ce{C2H5CHO}), the negligible differences throughout all stages can be attributed to the fact that these species have a very limited chemical reactions in the 2024 KIDA network with no grain surface reaction (either diffusive or non-diffusive) for propenal, just one pathway each for acetone and a couple for propanal.

In our model, most of the gas-phase COMs are majorly produced on the grain surface and subsequently desorbed to gas-phase through the chemical desorption following the prescription from \citet{riedel2023modelling}. Among the pathways only involving gas-phase species, dissociative recombination emerges as sole reaction mechanism. The destruction of gas-phase COMs is also mainly driven by a single mechanism of proton-transfer, where ions like \ce{H3+} and \ce{H3O+} act as primary proton donors.

Diffusive chemistry dominates over non-diffusive mechanisms for the gas-phase COMs for which the major formation pathway is successive hydrogenation on the grain-surface followed by reactive desorption in the final step, as the lone \ce{H} atom is light enough to efficiently diffuse at low temperatures of $\sim 10\text{ K}$. For instance, \ce{CH3OH} is most efficiently produced via a diffusive pathway through the hydrogenation of both the methoxy (\ce{CH3O}) and hydroxymethyl (\ce{CH2OH}) radicals, each contributing approximately 44\% to the overall formation. However, in the case of heavier COMs like ethanol (\ce{CH3CH2OH}), for which the major pathway ($\sim 55\%$) is
\vspace{-2em}
\begin{center}
    \[
        \ce{s-CH3 + s-CH2OH -> s-CH3CH2OH -> CH3CH2OH} \tag{R3} \label{reac:effnondiffgas1}
    \]
\end{center}
where both the reactants are heavy molecules, the non-diffusive pathway dominates. The prefix `s-' denotes the species is an ice and occurs on the grain surface. This trend can be seen for all COMs shown in Figure \ref{fig:majorcomsevolgas}. However, the relative contribution of each chemical process—gas-phase, grain-surface (either diffusive or non-diffusive)—is also time-dependent. For instance, acetaldehyde (\ce{CH3CHO}) forms primarily ($\sim$53\%) through a non-diffusive grain-surface reaction between \ce{s-CH3} and \ce{s-HCO} at the best-fitting time of 4 Myr. The formation of acetone (\ce{CH3COCH3}), methyl formate (\ce{HCOOCH3}), and methylamine (\ce{CH3NH2}) is notably dominated by gas-phase processes. Specifically, these molecules are primarily produced via dissociative recombination of their respective protonated precursors—\ce{C3H6OH+} for acetone (\(\sim 54\%\)), \ce{H5C2O2+} for methyl formate (\(\sim 80\%\)), and \ce{CH3NH3+} for methylamine (\(\sim 69\%\)) at the best-fitting time of 4 Myr. Finally, propanal (\ce{C2H5CHO}) is first formed on the grain surface through a non-diffusive reaction and subsequently desorbs into the gas phase via chemical desorption. Instead of being formed through hydrogenation, its formation occurs via the addition of atomic oxygen to propene (\ce{CH3CHCH2}). 

\subsubsection{On COMs in the ice-phase}
\begin{figure*}[htbp]
    \centering
    \includegraphics[width=\textwidth]{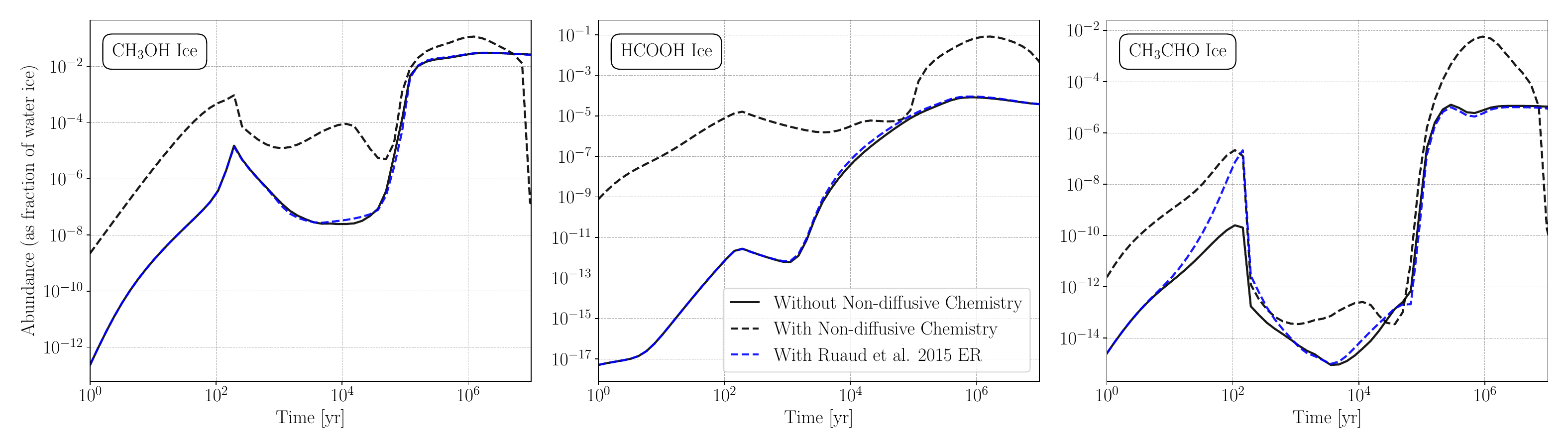}\\
    \includegraphics[width=0.66\textwidth]{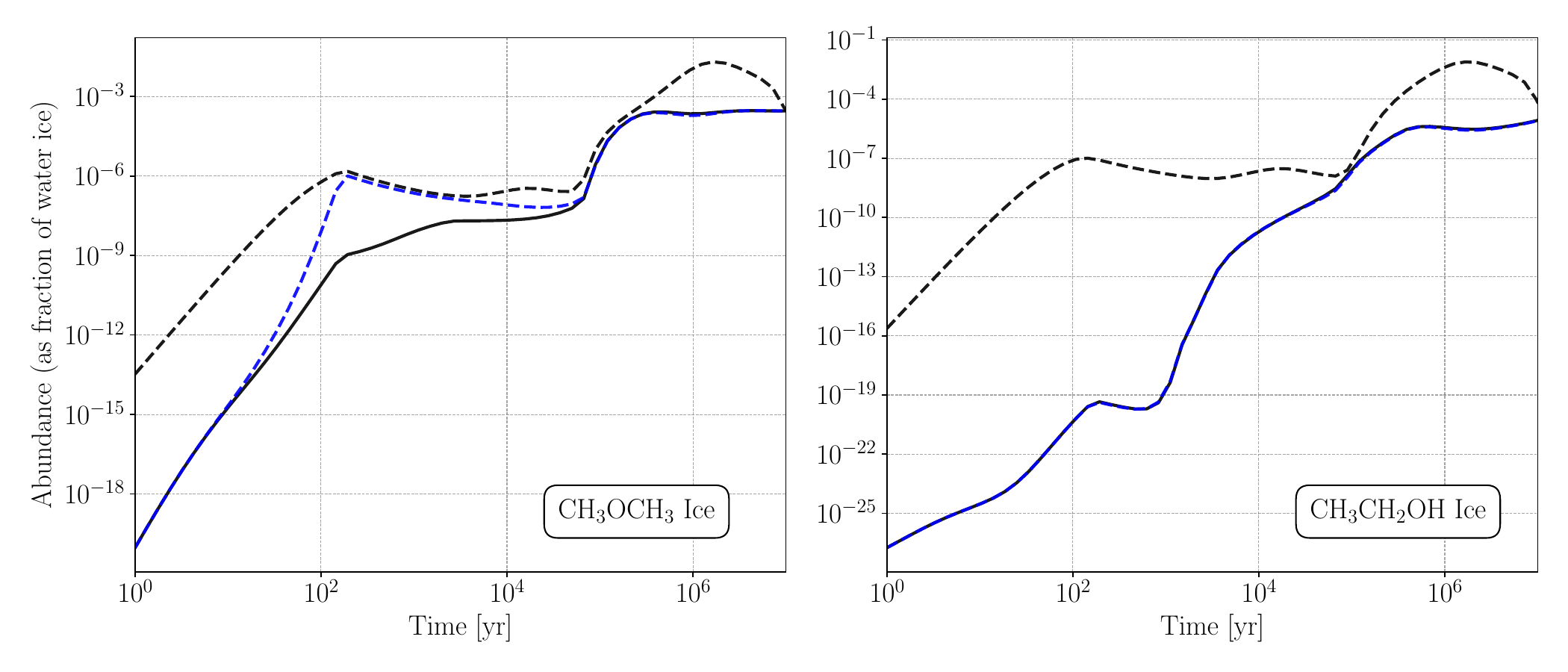}\\
    \includegraphics[width=0.66\textwidth]{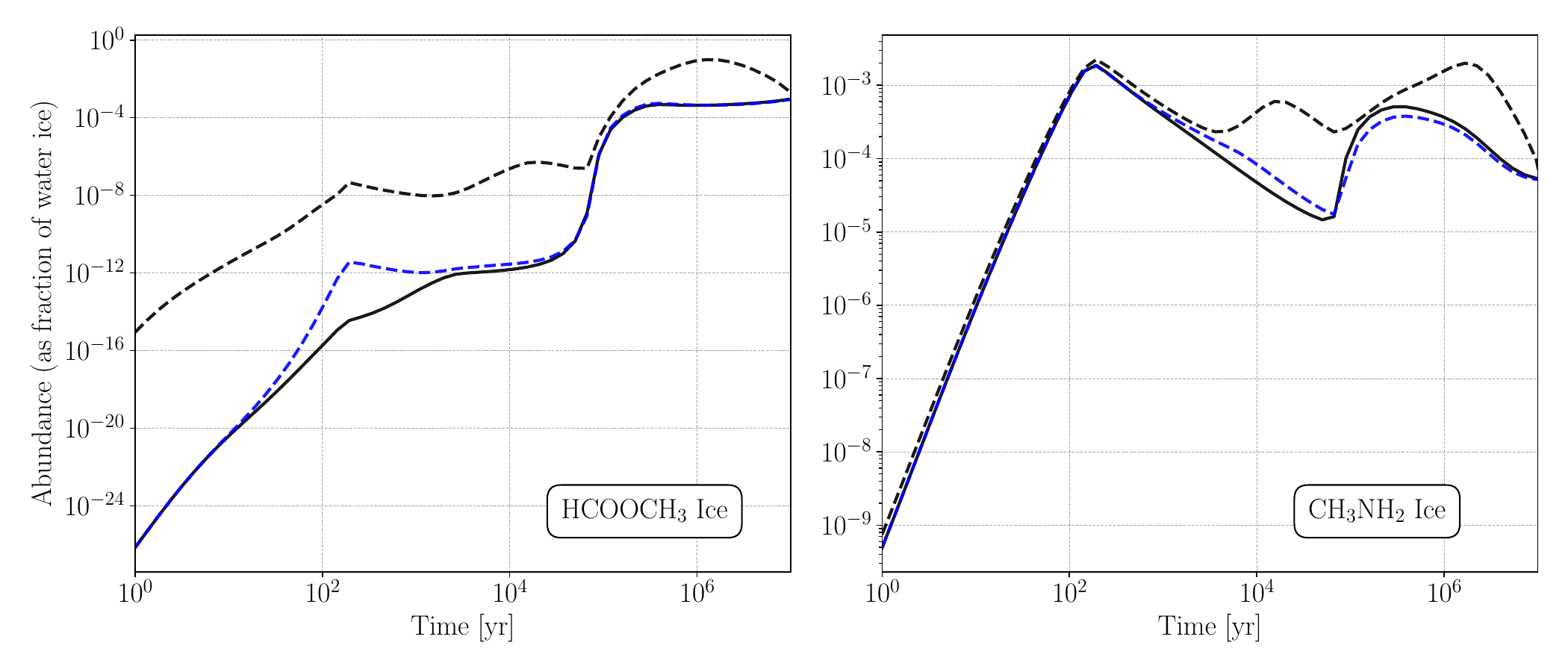}
    \caption{Abundances of major COM ices as a function of time with (model M3) and without (model M1) the inclusion of non-diffusive chemistry.}\label{fig:majorcomsevol}
\end{figure*}
We \textcolor{green}{concentrate} our discussion on a subset of the COMs listed in Table \ref{tab:species_data} as they have been detected in MYSOs, LYSOs and cometary bodies, and on a few oxygen-bearing COMs that have been detected in pre-stellar cores such as L1689B, L1544 and B1-b \citep{bacmann2012detection,cernicharo2012discovery,vastel2014origin}.  The evolution of the fractional abundances with respect to water ice for these are included through Figure \ref{fig:majorcomsevol}. 

To form methanol (\ce{s-CH3OH}) on the grain surface, without non-diffusive chemistry  there are two main pathways:
\vspace{-2em}
\begin{center}
    \[
        \ce{s-H + s-CH3O -> s-CH3OH} \tag{R4} \label{reac:effnondiffmeth1}
    \]
    \[
        \ce{s-H + s-CH2OH -> s-CH3OH} \tag{R5} \label{reac:effnondiffmeth2}
    \]
\end{center}

Reaction \ref{reac:effnondiffmeth1} represents the final step to form methanol via successive hydrogenation of \ce{s-CO}, whereas reaction \ref{reac:effnondiffmeth2} is hydrogenation of the hydroxymethyl radical, which itself originates through H-abstraction of methanol. With non-diffusive chemistry, the reaction \ce{s-OH + s-CH3 -> s-CH3OH} also becomes significant through the three-body non-diffusive reaction, but its contribution is still below 1\%. This is similar to the findings of \citet{jimenez2025modelling}, where this pathway was found to produce methanol only marginally, even by the models with non-diffusive chemistry. %This reaction
% \textcolor{red}{LM: Better word needed than pops}
%is enhanced through the three-body non-diffusive mechanism, whose precursors involve new non-diffusive 
% because of enhancements in the abundances of both reactants.  
Overall, the Eley-Rideal non-diffusive reactions \textcolor{green}{dominate all other non-diffusive pathways for methanol}. The destruction of \ce{s-CH3OH} is also greatly influenced by the inclusion of non-diffusive chemistry. In the diffusive only scenario, methanol is destroyed by reacting with atomic hydrogen leading to the formation of the hydroxy-methyl group (\ce{s-CH2OH}) or methoxide (\ce{s-CH3O}). Both of these processes are enhanced by three-body and Eley-Rideal non-diffusive mechanisms. Notably, non-diffusive chemistry is more efficient in destroying methanol on grain surfaces than photoprocesses.

For formic acid (\ce{s-HCOOH}), non-diffusive chemistry greatly boosts the main pathway \ce{s-H + s-HOCO -> s-HCOOH}, with \ce{s-OH + s-HCO -> s-HCOOH} also \textcolor{green}{making a} substantial contribution. The latter reaction proceeds through all non-diffusive mechanisms \textcolor{green}{(Eley-Rideal, three-body and photodissociation-induced non-diffusive mechanisms)} included in our model and is much more efficient than the diffusive counterpart. Turning on non-diffusive chemistry also enhances the rates of diffusive reactions, as more ices are available to react on the grain surfaces. For acetaldehyde (\ce{s-CH3CHO}), we have following pathways
\vspace{-2em}
\begin{center}
    \[
        \ce{s-CH3 + s-HCO -> s-CH3CHO} \tag{R6} \label{reac:effnondiffmeth5}
    \]
    \[
        \ce{s-H + s-CH3CO -> s-CH3CHO} \tag{R7} \label{reac:effnondiffmeth6}
    \]
\end{center}
We note that \ref{reac:effnondiffmeth5} is more efficient through the non-diffusive channel and \ref{reac:effnondiffmeth6} dominates for diffusive chemistry. The inclusion of non-diffusive chemistry also enhances destruction pathways for acetaldehyde, supplementing the conventional photoprocesses and diffusive reactions. We note that all non-diffusive mechanisms are more efficient than photoprocesses but less so than the diffusive counterparts.

The case of dimethyl ether (\ce{CH3OCH3}) is more interesting as it is produced through the following three-body excited non-diffusive reaction (represented as two successive two-body reactions)
\vspace{-2em}
\begin{center}
    \[
        \ce{s-CH^{\star}3 + s-H2CO -> s-CH3OCH2} \tag{R8} \label{reac:3bef1},
    \]
    \[
        \ce{s-H + s-CH3OCH2 -> s-CH3OCH3}
         \tag{R9} \label{reac:3bef2}
    \]
\end{center}
Reaction \ref{reac:3bef2} through this mechanism proceeds at a much higher rate than the diffusive counter-part due to the involvemnt of excited \ce{s-CH^{$\star$}3} ice in \ref{reac:3bef1}.
% \textcolor{green}{This is} the second most dominant pathway to produce \ce{CH3OCH3} after \ce{H + CH3OCH2 -> CH3OCH3}. 
The reaction \ce{s-CH3 + s-CH3O -> s-CH3OCH3} \textcolor{green}{is the other} major pathway to form dimethyl ether on the grain surface. Production of ethanol (\ce{s-CH3CH2OH}) is almost negligible without non-diffusive chemistry, but is amplified by the reaction of a more readily available hydroxy-methyl (\ce{s-CH2OH}) reacting with \ce{s-CH3} ice. Notably, turning on non-diffusive chemistry affects the diffusive reaction negatively in this case. Similarly, production of \ce{s-HCOOCH3} through the reaction of methoxide (\ce{s-CH3O}) with \ce{s-HCO} becomes much more feasible with non-diffusive chemistry, while suppressing the diffusive pathway. \textcolor{green}{While non-diffusive chemistry produces the smallest difference in abundances for methylamine (\ce{s-CH3NH2})}, a new pathway in the form of \ce{s-NH2 + s-CH3 -> s-CH3NH2} emerges as one of the major ways to form this nitrogen-bearing COM.

We \textcolor{green}{made the} same analysis with model M1 with ER chemistry from \citet{ruaud2015modelling} activated, and \textcolor{green}{found this was not as efficient as non-diffusive chemistry in producing the} COMs discussed here (Figure \ref{fig:majorcomsevol}).

\section{Conclusions}\label{sec:concl}
% Discuss everything mentioned in Wakelam2024 discussions
In this work, we presented a new and flexible astrochemical code which offers a large number of choices and options for exploring the effects of different chemical processes in varying physical conditions for interstellar medium. The key findings are summarized as below.

\begin{enumerate}
    \item Our benchmarking with well-established models, such as \textsc{Nautilus}, reveals excellent agreement between the two while also highlighting the key differences between the latest 2024 KIDA network and the previously released 2014 version. It also establishes the significance of reproducing identical results from identical initial conditions, implying robustness and stability of the numerical methods used in \textsc{Pegasis}. 
    \item The importance of grain surface chemistry in cold cores suggests that the three-phase prescription of \citetalias{Ruaud2016} should be able to capture the surface processes slightly better when compared to \citetalias{hasegawa1993three} \textcolor{red}{due to greater concentrations of species on the surface}.
    \item We find that current networks are still lacking in terms of unambiguously reproducing the gas-phase observations of TMC-1 (CP). This can be attributed to a multitude of factors discussed in \citet{wakelam20242024} regarding the uncertainties in the chemical network. We still choose the late time as the best-fitting time on the grounds that it provides the higher mean confidence.
    \item The significance of desorption cannot be understated when producing an estimate in chemical ages. Previous works like \citet{garrod2006gas} have emphasized on their importance and the process is sensitive to the reaction network, \textcolor{green}{as well as} the kinds of desorptive processes considered.
    \item The comparison with the ice abundances \textcolor{green}{observed toward} BG stars, MYSOs, LYSOs, and cometary bodies provides insights into how much the chemical inventory transforms as the cloud collapses to reach the more evolved protostellar stages.
    \item We explore the effects of all non-diffusive mechanisms included in our model on the abundances of COMs. We note that the major reactions are greatly enhanced after enabling non-diffusive chemistry, thereby increasing the share of COMs in both surface and bulk ice.
    \item For gas-phase COMs, non-diffusive mechanisms emerge as the dominant pathways for their formation on grain surfaces, followed by chemical desorption, especially when the involved reactants are heavier than atomic hydrogen that do not diffuse efficiently at the low temperatures typical of cold cores.
    \item Non-diffusive chemistry also affects the diffusive chemistry, \textcolor{green}{either} negatively and positively for different molecules. The Eley-Rideal non-diffusive mechanism and three-body non-diffusive reactions are consistently more efficient than photodissociation-induced non-diffusive reactions, all of which have higher rates than their diffusive counterparts across the COMs.
    % \item We find that the influence of non-diffusive chemistry on the three-phase prescription of \citet{hasegawa1993three} is \textcolor{red}{more prominent when it comes to the estimation of the best-fitting time, as it appears to compensate for the lack of lower surface concentrations than \citet{Ruaud2016}.}
    \item The more general treatment of Eley-Rideal chemistry through non-diffusive means as opposed to the carbon-specific treatment of the Eley-Rideal process in \citet{ruaud2015modelling} leads to a better estimation of COM abundances in cold cores.
    \item \textcolor{green}{The inclusion of non-diffusive chemistry alongside chemisorption makes \textsc{Pegasis} an extremely versatile astrochemical code, enabling simulations of diverse astrochemical environments with varying grain-surface chemistry and processes. These environments may correspond to cold cores (as in this work) or high-temperature regions (\(>100\text{ K}\)), which we will explore in a future study.}

\end{enumerate}

% We designed \textsc{Pegasis} to be easily extensible in terms of use, 

\begin{acknowledgements} 

L.M. expresses his sincere thanks to Valentine Wakelam for permitting the use of the \texttt{NAUTILUS} gas-grain three-phase chemical code in the past, for allowing a detailed benchmark comparison with \texttt{PEGASIS} in this work, and for encouraging the independent development of this new, fast, Python-based astrochemical code within his group. L.M. also thanks Ewine F. van Dishoeck for valuable discussions in the past related to the role of non-diffusive chemistry. L.M. acknowledges financial support from DAE and the DST-SERB research grant (MTR/2021/000864) from the Government of India for this work. This research was carried out in part at the Jet Propulsion Laboratory, which is operated for NASA by the California Institute of Technology. K.W. acknowledges the financial support from the NASA Emerging Worlds grant 18-EW-182-0083. We would like to thank the anonymous referee for constructive comments that helped improve the manuscript.

\end{acknowledgements}
\bibliography{bibliography}{}
\bibliographystyle{aa}
\newpage
\appendix
\section{Generation of Non-diffusive reactions}\label{app:gennondiff}
To generate the reaction set for each non-diffusive mechanism discussed in the
work, we start with grain-surface and bulk-ice reactions of the 2024 KIDA
network. To generate the set for photodissociation-induced non-diffusive
reactions, we identify all the species which are produced through
photodissociations by UV photons on grain surfaces and photodissociations by
cosmic-ray induced UV photons on grain surfaces. These are the \texttt{type
    17, 18, 19, 20} in the reaction network \citep{wakelam20242024}. Once we have
these \textit{photoproducts}, we move to \texttt{type 14} which represents all
diffusive reactions, to identify the subset of reactions where the
photoproducts participate as reactants, giving us the required reaction set. In
the case of non-diffusive Eley-Rideal reactions, similar process is performed,
except this time, we take products from \texttt{type 99}, which represents
adsorption reactions on grains. In the subset of diffusive reactions where
these products occur as reactants, it ensures that one reactant has already
accreted on the grain-surface and it immediately encounters the other reactant
there. Further, we ensure that the complete reaction is a surface-only process
with no bulk-ice species involved, as is the case with E-R. Finally, to
generate the non-diffusive three-body reactions, we need to select the subset
of diffusive reactions for both grain-surface and grain-mantle where both
involved reactants occur as products of diffusive reactions itself. To this
end, we take products from \texttt{type 14}, and then proceed similar to
previous cases to generate the reaction set for this mechansim. The following two
grain-surface reactions are added separately for the three-body excited formation mechanism \citep{jin2020formation}
\vspace{-2em}
\begin{center}
    \[
        \ce{s-CH^{\star}3 + s-CO -> s-CH3CO} \tag{AR1} \label{reac:appbef1}
    \]
    \[
        \ce{s-H + s-CH3CO -> s-CH3CHO} \tag{AR2} \label{reac:appbef1ex}
    \]
\end{center}
and
\vspace{-2em}
\begin{center}
    \[
        \ce{s-CH^{\star}3 + s-H2CO -> s-CH3OCH2} \tag{AR3} \label{reac:appbef2}
    \]
    \[
        \ce{s-H + s-CH3OCH2 -> s-CH3OCH3} \tag{AR4} \label{reac:appbef2ex}
    \]
\end{center} 
under the constraint that $\ce{s-CH^{\star}3}$ ice is formed by the following reaction
\vspace{-2em}
\begin{center}
    \[
        \ce{s-H + s-CH2 -> s-CH^{\star}3} \tag{AR5} \label{reac:appbef3}
    \]
\end{center}
\citet{jin2020formation} discuss an additional reaction that undergoes a non-diffusive process via the formation of an excited intermediate. However, this reaction was not included because one of its participants (\ce{s-CH3OCO}) is not present in the 2024 KIDA network. 
\section{Benchmarking \textsc{Pegasis}}\label{app:bench}
\begin{figure*}[htbp]
    \centering
    \includegraphics[width=0.75\textwidth]{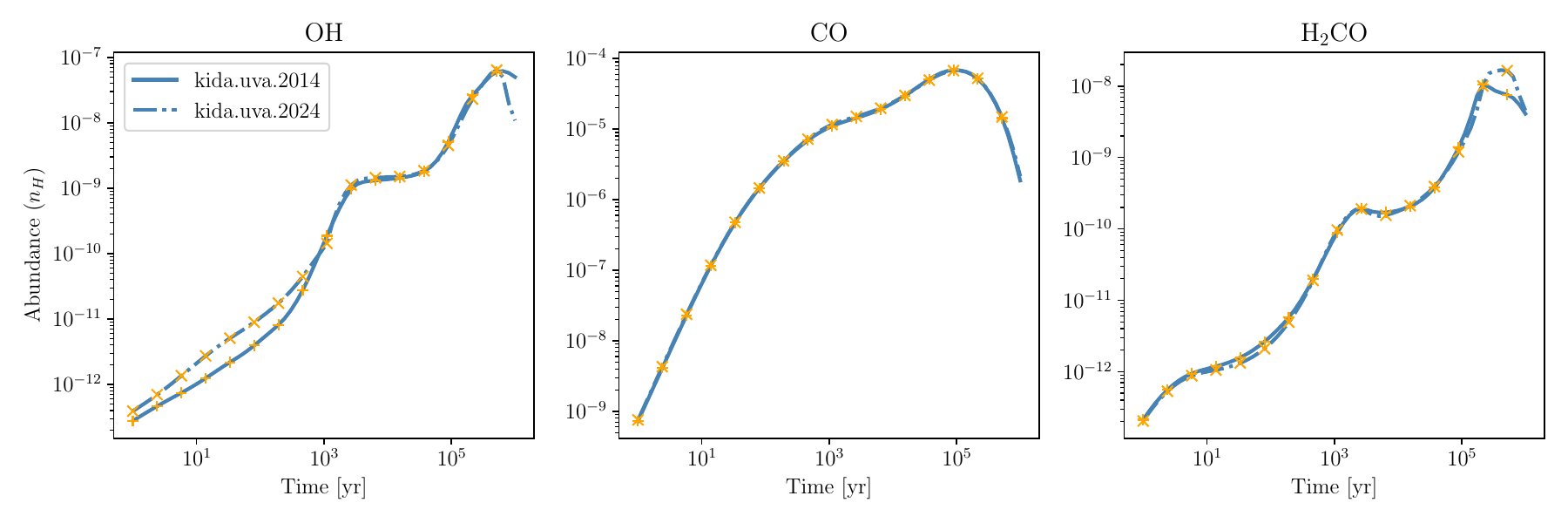}\\
    \includegraphics[width=0.75\textwidth]{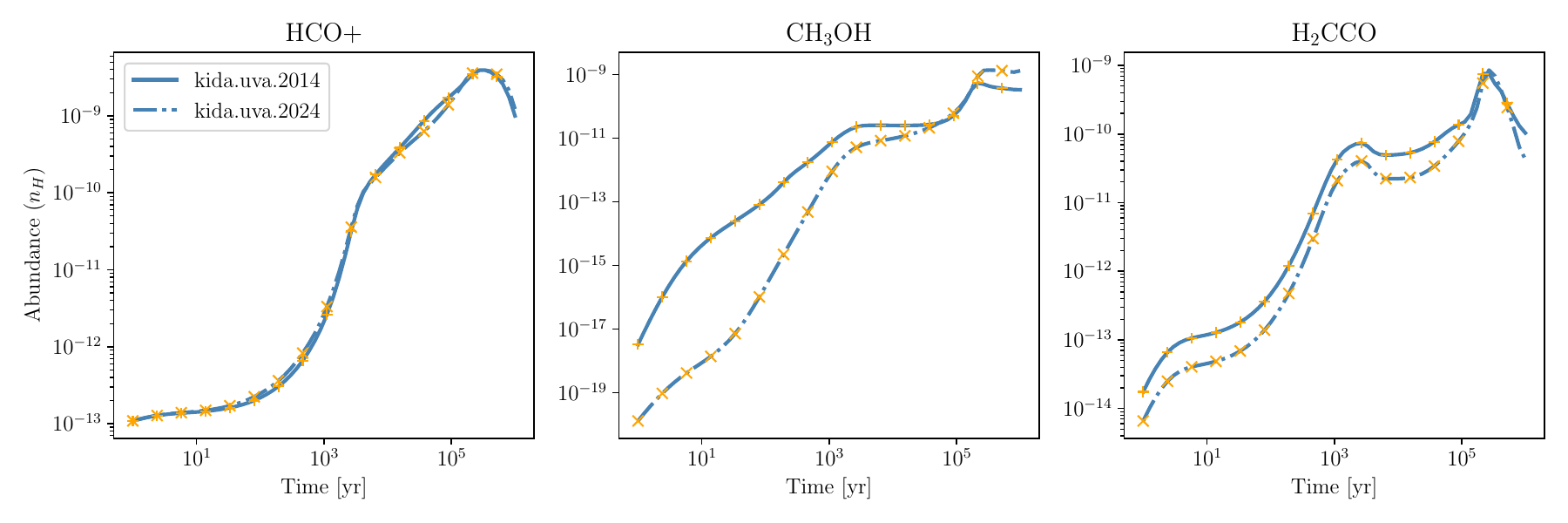}\\
    \includegraphics[width=\textwidth]{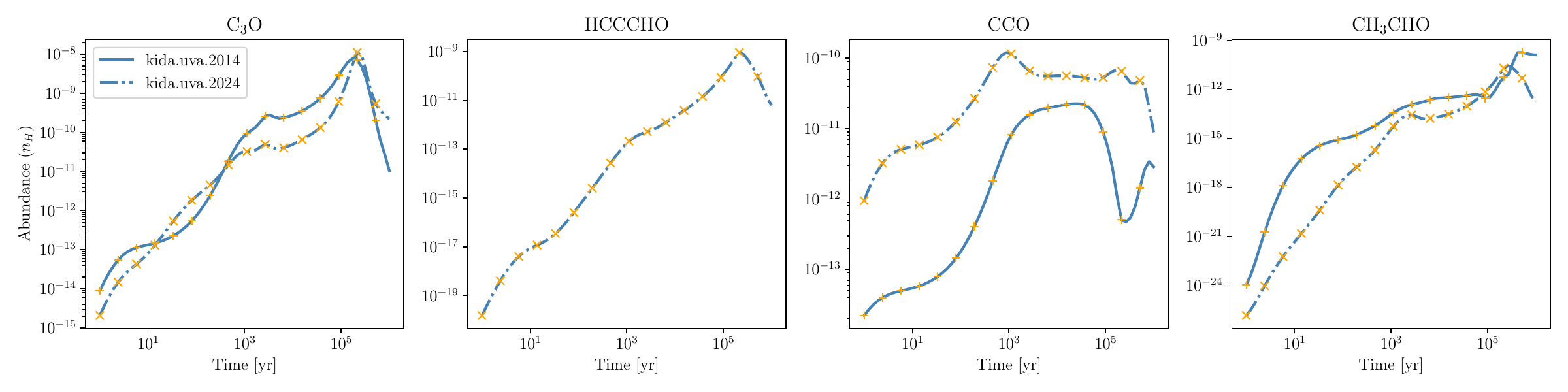}
    \caption{Comparison of selected oxygen-bearing species in gas-phase detected in TMC-1 molecular cloud. Both \textsc{Pegasis} (blue lines) and \textsc{Nautilus} (orange markers) have excellent agreement for both the networks (2014 is represented by dashdotted lines and 2024 represented by solid lines).}
    \label{fig:oxygen-bearings}
\end{figure*}

\begin{figure*}[htbp]
    \centering
    \includegraphics[width=\textwidth]{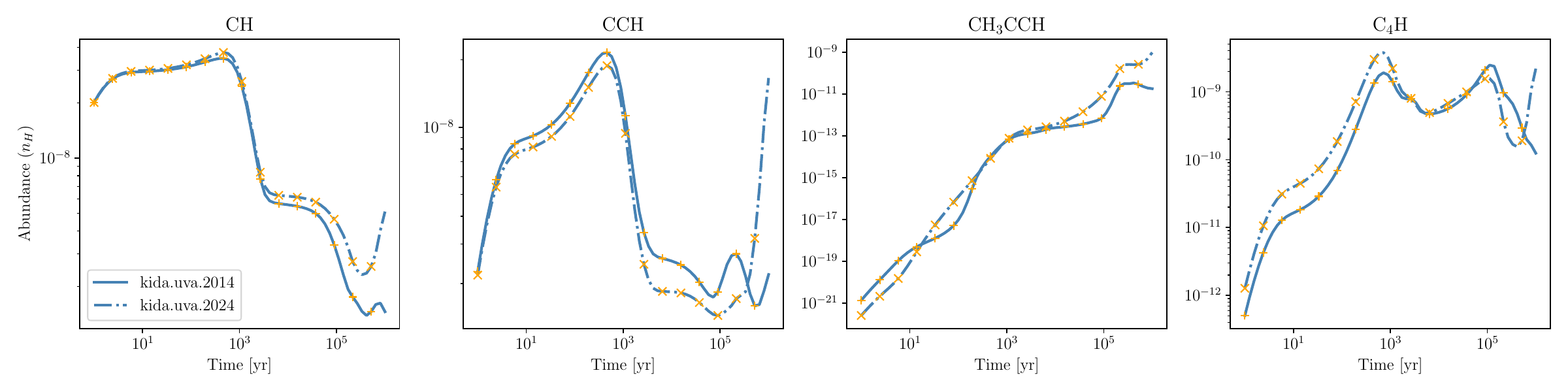}\\
    \includegraphics[width=\textwidth]{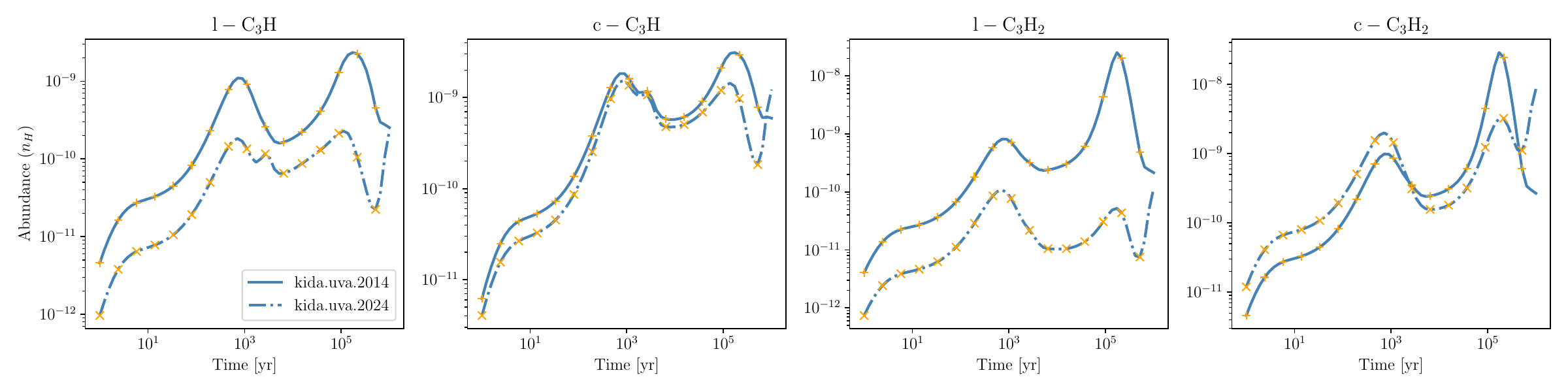}\\
    \includegraphics[width=\textwidth]{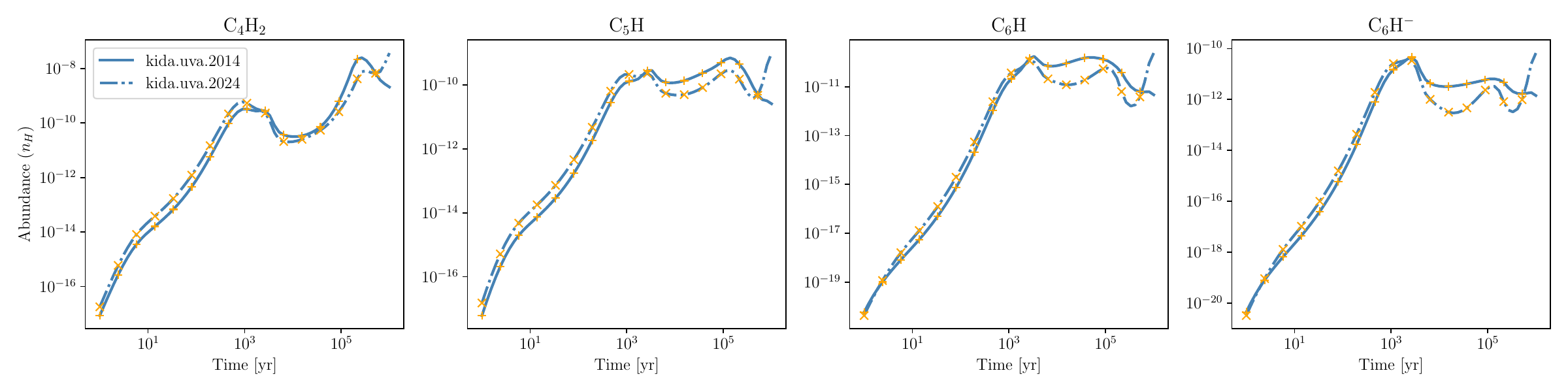}\\
    \includegraphics[width=\textwidth]{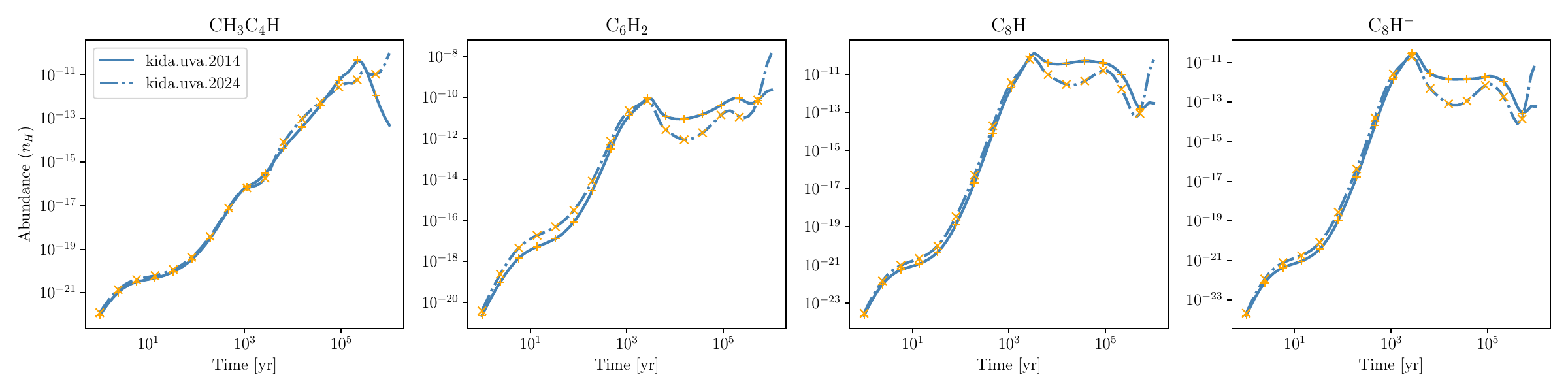}
    \caption{Same as Figure \ref{fig:oxygen-bearings}, but for hydrocarbons.}
    \label{fig:carbon-bearings}
\end{figure*}

\begin{figure*}[htbp]
    \centering
    \includegraphics[width=\textwidth]{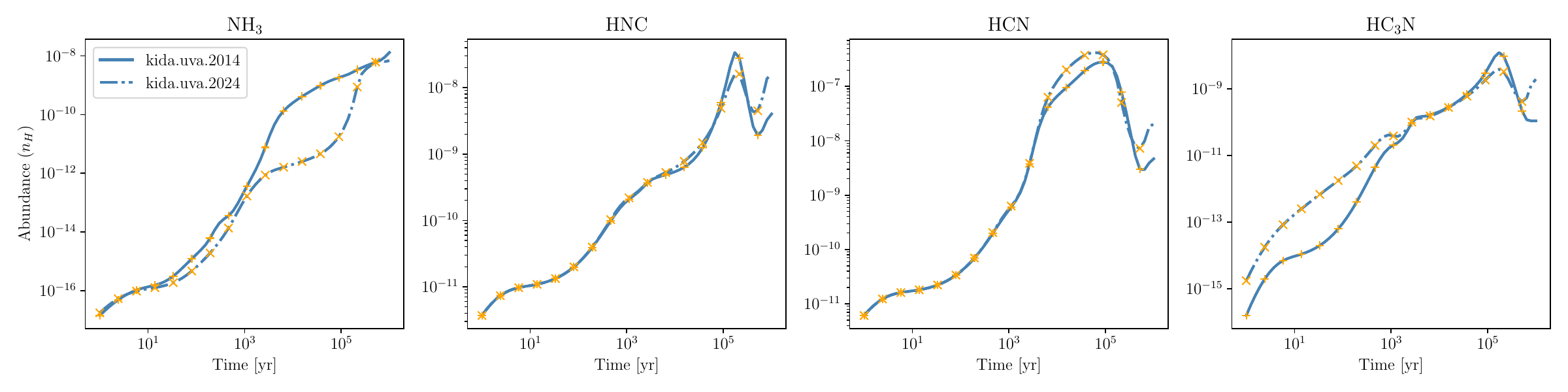}\\
    \includegraphics[width=\textwidth]{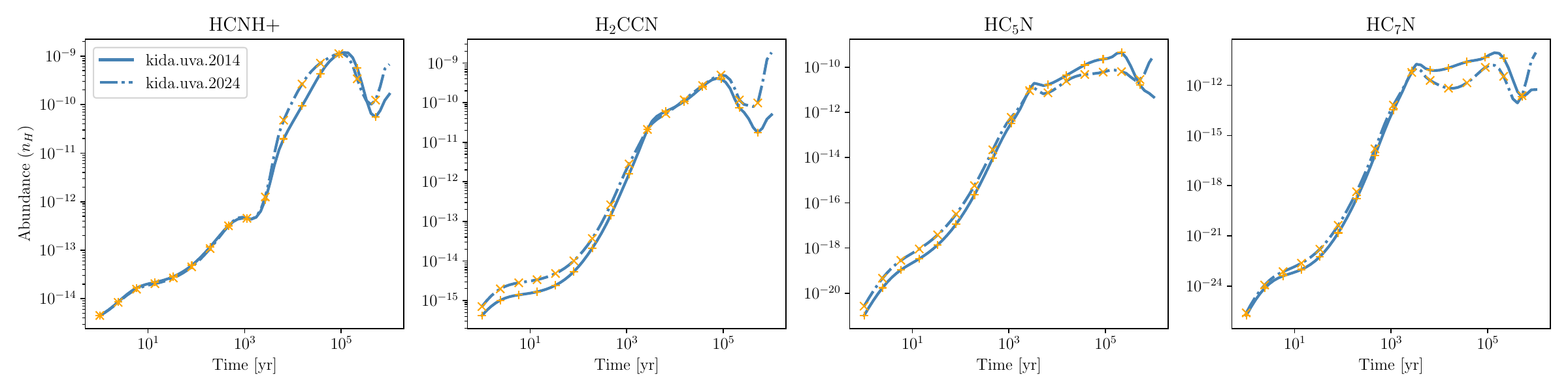}\\
    \includegraphics[width=\textwidth]{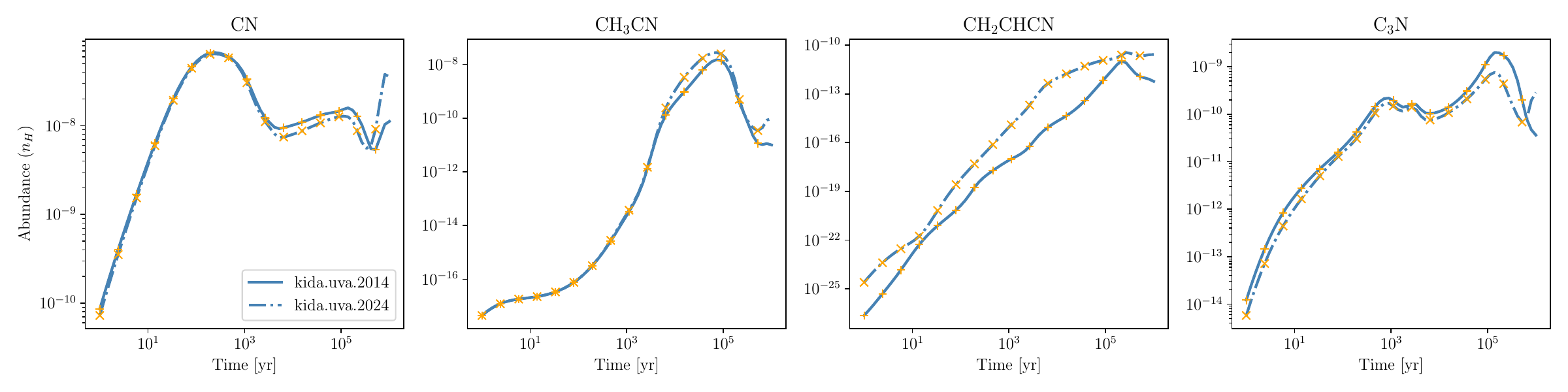}\\
    \includegraphics[width=\textwidth]{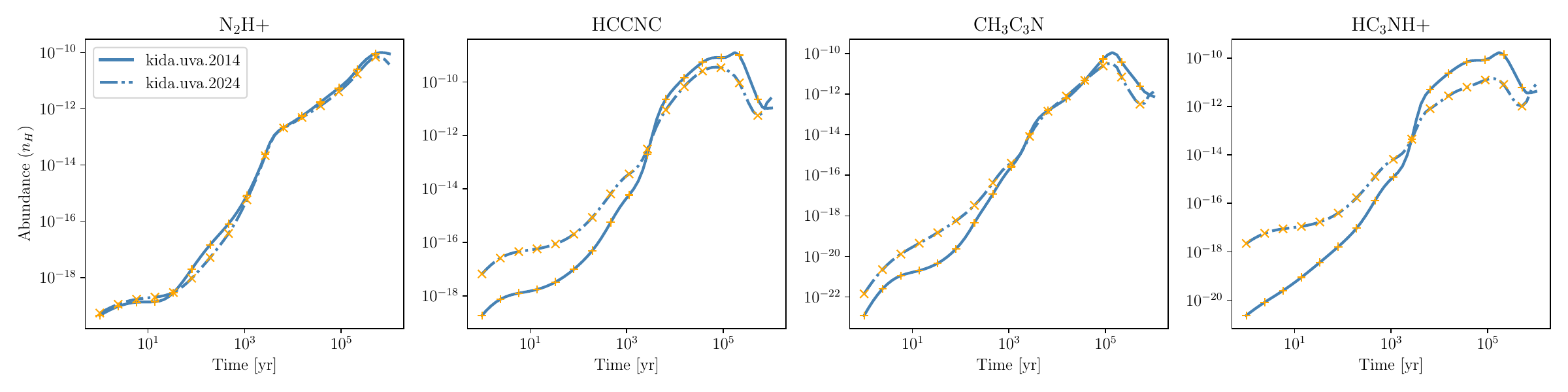}\\
    \includegraphics[width=\textwidth]{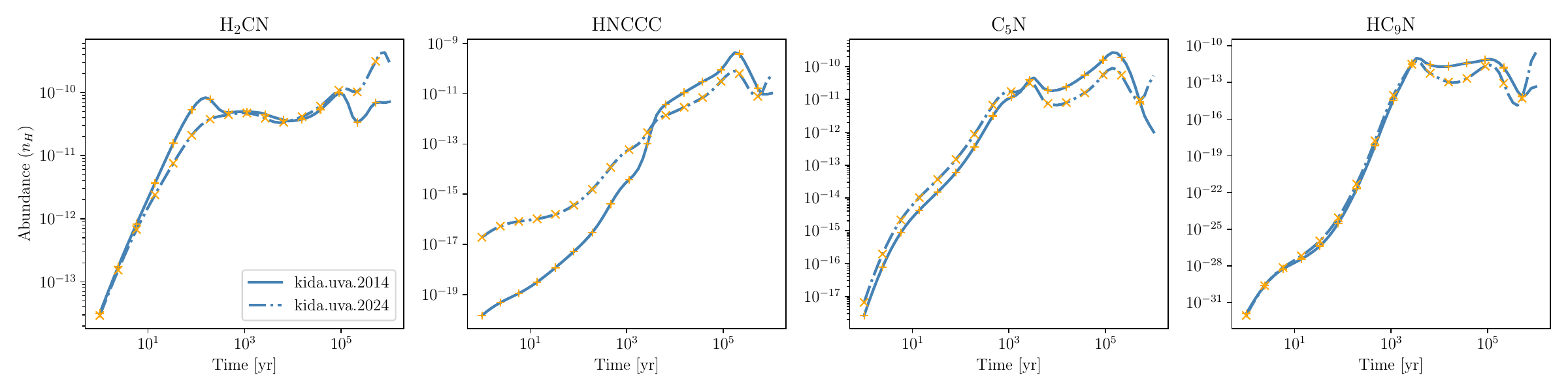}
    \caption{Same as Figure \ref{fig:oxygen-bearings}, but for nitrogen-bearing species.}
    \label{fig:nitrogen-bearings}
\end{figure*}

\begin{figure*}[htbp]
    \centering
    \includegraphics[width=0.5\textwidth]{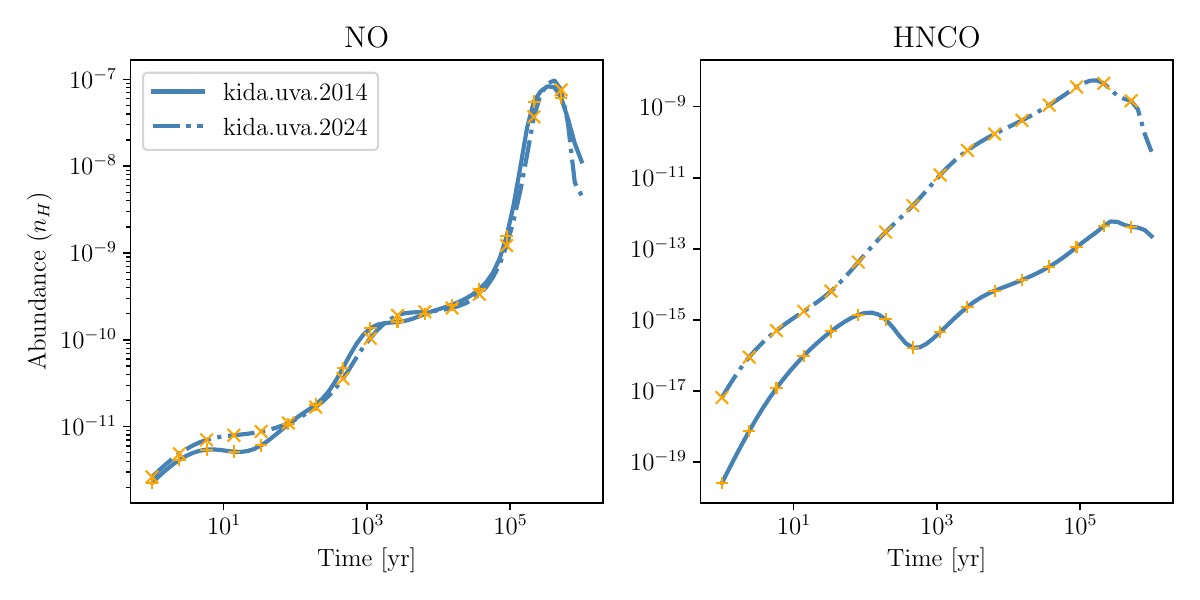}
    \caption{Same as Figure \ref{fig:oxygen-bearings}, but for NO-bearing species.}
    \label{fig:no-bearing}
\end{figure*}

\begin{figure*}[htbp]
    \centering
    \includegraphics[width=\textwidth]{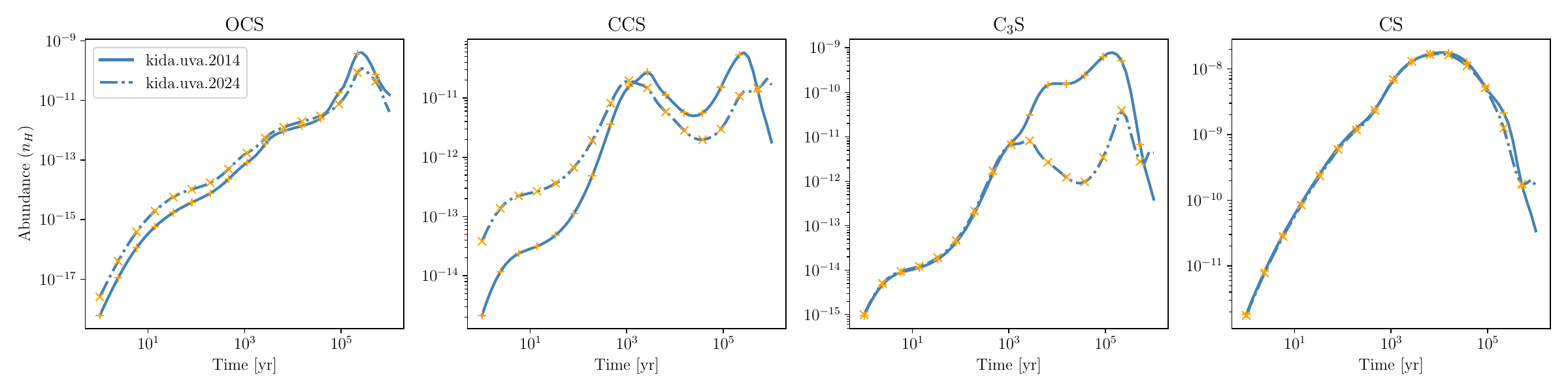}\\
    \includegraphics[width=0.75\textwidth]{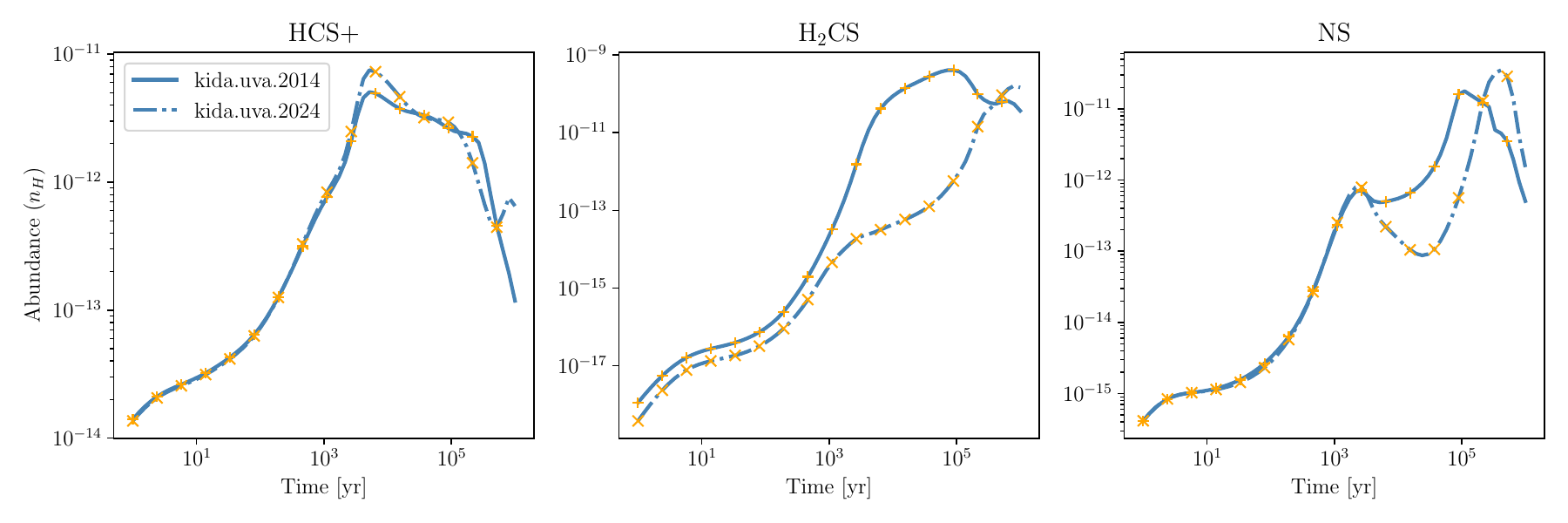}\\
    \includegraphics[width=0.5\textwidth]{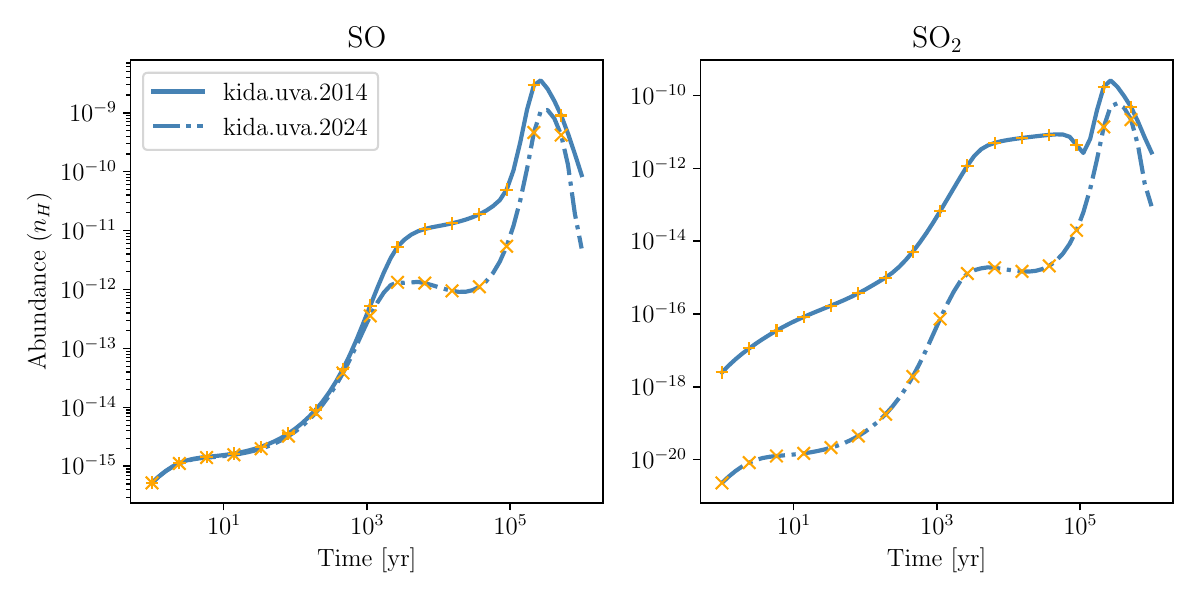}
    \caption{Same as Figure \ref{fig:oxygen-bearings}, but for sulfur-bearing species.}
    \label{fig:sulfur-bearings}
\end{figure*}

We depict time evolution of select species detected in TMC-1 (CP) using the both 2014 and 2024 KIDA networks with \textsc{Pegasis} and \textsc{Nautilus}. 
% \textcolor{red}{LM: We don't need add observation here: We also plot \(\pm 0.3\) dex around the observed abundance. Most of these signify an early time, but as discusses, a change in switches related to desorption and other model parameters can influence such results.}
The oxygen-bearing species are shown in Figure \ref{fig:oxygen-bearings}, with hydrocarbons in Figure \ref{fig:carbon-bearings}, nitrogen-bearing in Figures \ref{fig:nitrogen-bearings}, \ref{fig:no-bearing} and sulfur-bearing species follow in Figure \ref{fig:sulfur-bearings}. For visual clarity, the \textsc{Nautilus} results were over-plotted over \textsc{Pegasis} results with a slightly larger line-width and lower opacity. \textcolor{green}{As is evident from the figures, \textsc{Pegasis} and \textsc{Nautilus} have excellent agreement, with the plots of both models exhibiting high co-incidence.}
% \section{Modelling prestellar core L1544}
% \lipsum[45-48]

\end{document}